\documentstyle[aps,amssymb,twocolumn]{revtex}
\begin{document}
\draft
\title{Theory of the first-order isostructural valence phase transitions in mixed
valence compounds YbIn$_x$Ag$_{1-x}$Cu$_4$.}
\author{A.V.Goltsev}
\address{Ioffe Physical-Technical Institute, 194021 St.Petersburg, Russia}
\author{G.Bruls}
\address{Physikalisches Institut, Universit\"{a}t Frankfurt, 60054 Frankfurt,\\
Germany}
\date{\today}
\maketitle

\begin{abstract}
For describing the first-order isostructural valence phase transition in
mixed valence compounds we develop a new approach based on the lattice
Anderson model. We take into account the Coulomb interaction between
localized $f$ and conduction band electrons and two mechanisms of
electron-lattice coupling. One is related to the volume dependence of the
hybridization. The other is related to local deformations produced by $f$
shell size fluctuations accompanying valence fluctuations. The large $f$%
-state degeneracy allows us to use the $1/N$ expansion method. Within the
model we develop a mean-field theory for the first-order valence phase
transition in YbInCu$_4$. It is shown that the Coulomb interaction enhances
the exchange interaction between $f$ and conduction band electron spins and
is the driving force of the phase transition. A comparison between the
theoretical calculations and experimental measurements of the valence
change, susceptibility, specific heat, entropy, elastic constants and volume
change in YbInCu$_4$ and YbAgCu$_4$ are presented, and a good quantitative
agreement is found. On the basis of the model we describe the evolution from
the first-order valence phase transition to the continuous transition into
the heavy-fermion ground state in the series of compounds YbIn$_{1-x}$Ag$_x$%
Cu$_4$. The effect of pressure on physical properties of YbInCu$_4$ is
studied and the $H-T$ phase diagram is found.
\end{abstract}


\narrowtext

\section{Introduction}

The first-order isostuctural valence phase transition observed in rare earth
intermetallic compounds is one of the most striking and interesting
phenomena in strongly correlated electron systems. For the first time the
transition was discovered in elemental Ce which at applied pressure shows a $%
\alpha \rightarrow \gamma $ transition (for a review, see \cite
{KandG,Lawrence}). Other compounds demonstrating the transition are samarium
monochalcogenides, \cite{Smirnov,Jayaraman} a few Eu-based compounds, \cite
{Croft,Michels} CeNi$_{x}$Co$_{1-x}$Sn, \cite{Adroja} and YbInCu$_{4}.$\cite
{FelnerNowik86,FelnerNowik87,FelnerNowik88} The latter compound has
attracted much recent attention because YbInCu$_{4\text{ }}$ demonstrates a
first-order isostuctural valence phase transition at ambient pressure. The
remarkable peculiarity of the transition lies in the small change of valence
of the Yb ions from nearly +3 at high temperatures ($T>$50 K) to about
2.8-2.9 below the critical temperature $T_{v}=42$ K. The volume expansion of
YbInCu$_{4}$ taking place at $T=T_{v}$ is also very small, about +0.5\%,
whereas in Ce the volume change achieves -15\%. Despite the small valence
and volume changes at $T=T_{v}$, many physical properties of YbInCu$_{4}$
are changed dramatically. Already first magnetic measurements\cite
{FelnerNowik87,FelnerNowik88} showed that at the transition the
susceptibility drops abruptly, and the Curie-Weiss temperature dependence
above $T_{v}$ is replaced by an almost temperature independent behavior
which is typical for a nonmagnetic metal with an enhanced Pauli paramagnetic
susceptibility. While the Curie-Weiss behavior at high temperatures can be
easily related to the magnetic moment $J=7/2$ of trivalent Yb ions, the
formation of the Pauli paramagnetic state is less clear, taking into account
that the valence change $\Delta N_{f}$ at $T_{v}$ is estimated to be only
about 0.1 as it was revealed by x-ray absorption measurements at the Yb $%
L_{3}$ edge.\cite{FelnerNowik87,Cornelius97} From the same point of view the
results of Hall effect measurements \cite{Cornelius97,Figuer} look very
amazing, since they give evidence for a very large change of the charge
carrier concentration from 0.07 per formula unit at high temperatures to 2.2
per formula unit below $T_{v}$. In the high temperature phase the Hall
measurements are consistent with the band calculations \cite
{TakegaharaKasuya} which showed that in this phase YbInCu$_{4}$ has a
semimetal band structure similar to LuInCu$_{4}$. Transport measurements in
YbInCu$_{4}$ also shows that the compound is semimetalic above $T_{v}$.\cite
{Nakamura94} Resistance measurements revealed a large and abrupt drop of
electrical resistivity when decreasing the temperature below $T_{v}$.
Investigations of Cu nuclear quadrupolar resonance \cite
{Nakamura90,Graham,Nakamura96} (NQR) and Cu and In Knight shift \cite
{Nakamura98} in YbInCu$_{4}$ show that at $T>T_{v}$ the 4$f$ electrons are
localized and interact weakly with the conduction band electrons while at $%
T<T_{v}$ a Fermi-liquid state with delocalized Yb 4$f$ electrons is formed.
The measurements revealed occurrence of a strong hybridization between Yb 4$%
f $ electron states and conduction band states in the Fermi-liquid state.
This result is in agreement with M\"{o}ssbauer experiments \cite
{FelnerNowik87} that also revealed a nonmagnetic state of the Yb ions below $%
T_{v}$. Additional evidence for the fact comes from the temperature behavior
of the entropy ($S$) inferred from specific heat measurements.\cite{Sarrao98}
Above $T_{v}$ the entropy is large and can be explained by a large
contribution from the $f$ electrons. At $T_{v}$ the entropy drops
significantly and with decreasing temperature it tends to zero in the limit $%
T\rightarrow 0$. It means that at $T=0$ the ground state of the electron
system consisting of conduction band and $f$ electrons, is a singlet state
which in accordance with magnetic, NQR and transport measurements is a
Fermi-liquid state.

Summarizing, one can conclude that the main mystery of YbInCu$_4$ is in the
following. In spite of a small valence change at the phase transition ($%
\Delta N_f\sim 0.1)$ from which one could expect that $f$ electrons would
save their localized character, the thermal, magnetic, NQR and transport
measurements mentioned above clearly demonstrate that below $T_v$ the $f$
electrons loose their localized character and form a nonmagnetic
Fermi-liquid state together with the conduction electrons .

Taking into account the nonmagnetic character of the Yb ions at $T<T_{v}$ ,
the enhanced values of the zero temperature susceptibility $\chi (0)$ and
linear specific heat coefficient $\gamma $, Felner et al.\cite{FelnerNowik87}
came to the conclusion that YbInCu$_{4\text{ }}$ is a ``light heavy-fermion
system''.

The uniqueness of the first-order valence phase transition in YbInCu$_{4%
\text{ }}$becomes especially clear when we compare the compound with other
isostructural Yb compounds, namely, YbXCu$_{4\text{ }}$with X=Ag, Au, Cd,
Mg, Tl, and Zn. A detailed study and comparison of the compounds have
recently been performed by Sarrao et al. \cite{Sarrao98pr} Though the
compounds have the same face-centered-cubic crystal structure, their
physical properties are very different. Among the compounds YbAgCu$_4$ is of
special interest. It is a typical heavy-fermion (HF) compound with no
magnetic order. A substitution of Ag by In results in crossover from the HF
behavior to the first-order isostructural valence phase transition, as it
has been found when investigating the series of compounds YbIn$_{1-x}$Ag$_x$%
Cu$_4.$ \cite{Pillmayr,Sarrao96,Cornelius97}

In order to explain the valence phase transition in Ce several models have
been proposed. In the framework of promotional models\cite{F-K,R-F} it is
assumed that there is an explicit change of the electronic configuration of
Ce ions from 4$f^1$ to 4$f^0$, and $f$ electrons transfer into the
conduction band. A Mott transition picture considers the valence transition
as a localization-delocalization transition in the system of $f$ electrons.%
\cite{Johansson} The Kondo volume-collapse (KVC) model\cite{Allen82,Allen92}
explains the valence phase transition by a strong volume dependence of the
Kondo temperature $T_k$, neglecting correlation effects between the Kondo
effect on different $f$ ions. Each of the models has advantages and
shortcomings. The shortcomings becomes especially contrasting when we try to
use the models for explaining an evolution of the valence phase transition
in the series of compounds YbIn$_{1-x}$Ag$_x$Cu$_4.$

The promotional models demand the valence change to be about 1, while in YbIn%
$_{1-x}$Ag$_{x}$Cu$_{4}$ the change is only about 0.1. This disagreement
does not allow within the models to explain quantitatively, using realistic
physical parameters, magnetic and other properties of the Yb compounds in
the low temperature phase. A recent attempt\cite{Zlatic} to correct the
deficiency of the Falicov-Kimbal model is based on the assumption that only
a fraction of Yb ions are active and other ones are strongly hybridized.
However, the assumption has no experimental confirmation. It contradicts ,
for example, recent thermoanalytical investigations \cite{Fischbach} of the
system Yb-In-Cu which showed that by choosing certain starting compositions
one can obtain such YbInCu$_{4}$-samples which have $T_{v}\approx 40$ K, an
improved site order and a reduced probability of Yb ions to occupy In-sites.
The Falicov-Kimbal model assumes that the 4$f^{14}$configuration is the
ground state of Yb ions and 4$f^{13}$ configuration is the first excited
state. This assumption contradicts with a large number of experimental data
that give direct and indirect evidences for the fact that the 4$f^{13}$
state is the ground state, and the 4$f^{13}$ configuration is the first
excited state (see below Sec.II). The promotional model actually neglects a
hybridization between $f$ states and conduction band states. It contradicts
with experimental data on NQR\cite{Nakamura90,Graham,Nakamura96} and the
Knight shift measurements \cite{Nakamura98} mentioned above. Moreover, on
the basis of the models it is impossible to explain the development of a
continuous transition into the HF ground state in YbIn$_{1-x}$Ag$_{x}$Cu$%
_{4} $ at $x>0.2$ \cite{Sarrao96}, as the Kondo effect taking place in the
compounds, can not be described in the framework of the model.

In terms of the Mott transition picture\cite{Johansson} at $T<T_v$ the $f$
band width $W$ has to become larger than the on-site Coulomb repulsion
energy $U$ between $f$ electrons. It is possible if the volume of the
lattice shrinks below $T_v$. Such a situation takes place in Ce, but not in
YbInCu$_4 $ . The volume of YbInCu$_4$ increases by 0.5\%, which evidences
against the Mott transition picture.

It should be mentioned that both the promotional models and Mott transition
picture have a charge fluctuation energy as the energy scale. Usually a
charge fluctuation energy is too large to provide a convincing explanation
of the small energy scales observed in YbIn$_{1-x}$Ag$_x$Cu$_4$.

The KVC model faces a serious problem when explaining pressure dependence of
the zero temperature susceptibility in YbInCu$_{4}$. The model needs the
Gr\"{u}neisen parameter $\Gamma $ to be negative and equal to $-3000$, an
unphysical value, while the experiment\cite{Sarrao98} gives $\Gamma \approx
-31.$ The negative sign of $\Gamma $ also constitutes a challenge to the
model because within the usual mechanism of the volume dependence of the
Kondo temperature due to the volume dependence of the hybridization, $\Gamma
$ should be positive. Moreover, the KVC model neglects correlations, or in
other words, coherence in the Kondo screening of nearby Yb ions in the low
temperature phase. However, it is the latter effect that results in the
formation of the heavy fermion state in a heavy fermion compound such as
YbAgCu$_{4}$(see, for example, reviews \cite{NewsRead,Schlottmann}).

In the present paper we propose a new theoretical approach that allows us
both to obtain quantitative agrement with various experimental data and to
avoid contradictions and deficiencies of the previous approaches. Our
approach is based on the conclusion obtained from the experimental data
mentioned above that the low temperature phase of YbIn$_{1-x}$Ag$_{x}$Cu$%
_{4} $ is the HF state. From this point of view the first-order valence
phase transition is a normal metal (semimetal)-HF metal transition. Besides,
it is supposed that at $T>T_{v}$ in the normal state the $f$ electrons are
localized on the $f$ shells and produce the magnetic moments $J=7/2$ of the
Yb ions. In the HF state, i.e. at $T<T_{v}$, $f$ states are strongly
hybridized with conduction band states, and as a result, quasiparticles near
the Fermi surface have a hybrid character and an enhanced mass. Such a
physical picture can be successfully described on the basis of the lattice
Anderson model. However, it is important to note that the conventional
lattice Anderson model allows to study only a continuous transition from the
normal state with the incoherent Kondo scattering into the heavy-fermion
state with the coherent Kondo effect, as it occurs in heavy-fermion
compounds.\cite{NewsRead} In the present paper we extend the model in the
following way. At first, we take into account a Coulomb repulsion between $f$
and conduction band electrons. Second, we take into account two mechanisms
of the electron-lattice coupling. The first mechanism is related to the
volume dependence of the hybridization between $f$ and conduction band
states due to a change of the overlapping between wave functions of the
conduction band and $f$ electron states. It is the conventional mechanism of
the electron-lattice interaction in HF compounds, that has been used
specifically in the KVC model. However, it is necessary to note that valence
fluctuations of the Yb ions are accompanied by fluctuations of the $f$ shell
size, which results in local lattice deformations. It is the second ($f$%
-shell-size-fluctuation) mechanism of the electron lattice coupling
considered in our approach. If one neglects crystal field splitting, then
the $f$ level has the degeneracy $N=2J+1=8$. Such a large degeneracy enables
us to use the $1/N$ expansion technique for attacking the extended lattice
Anderson model. In the present work we shall calculate the free energy in
the mean-field approximation, that corresponds to the leading order in $1/N$%
. A calculation of $1/N$ corrections is beyond the scope of our paper.

Our paper has the following structure. In Sec. II we introduce the extended
lattice Anderson model. The mean-field approach for studying the model is
developed in Sec. III. The electron-lattice coupling and volume change are
considered in Sec. IV. In the next Sec. V we investigate magnetic properties
of the system and magnetic field effect. Results of numerical calculations
based on the mean field approach are discussed in Sec. VI where we compare
the theoretical results on temperature and pressure dependences of thermal,
magnetic and elastic properties of YbInCu$_{4}$ and YbAgCu$_{4}$ with
experimental measurements. In Sec.VII we will discuss the role of
fluctuations around the mean-field solution in low and high temperature
phases and make conclusions..

\section{Model}

A number of experimental data give direct evidences for the fact that Yb
ions in YbInCu$_{4}$ fluctuate between the magnetic ground state Yb$^{+3}$
having the configuration 4$f^{13}$ with one hole in the $f$ shell and total
magnetic moment $j$=7/2, and the excited nonmagnetic state Yb$^{+2}$ having
a completely filled $f$ shell with the configuration 4$f^{14}$. One can
mention the M\"{o}ssbauer spectroscopy\cite{FelnerNowik87} and the
photoemission spectra (PES) measurements\cite{Ogawa}. Moreover, according to
the PES measurements, neither 4$f^{12}$ nor 4$f^{13}$ $^{2}F_{5/2}$
contribute to low energy properties. Therefore, one can neglect transitions
into these states.

For describing the system it is convenient to use a hole representation.
Valence fluctuations of Yb ions can be described in terms of hole
transitions between $f$ shells and the hole conduction band. The picture of
behavior of an 4$f$ ion in a metal was developed by Hirst \cite{Hirst}. It
underlies the theoretical model we will use in the paper. The system of
interacting $f$ holes and conduction band holes can be described using the
lattice Anderson Hamiltonian in the slave boson representation: \cite
{NewsRead,Coleman83,ReadNews83}\
\begin{eqnarray}
H_{A} &=&\sum_{a{\bf k}}\varepsilon _{{\bf k}}c_{\alpha {\bf k}%
}^{+}c_{\alpha {\bf k}}+\sum_{\alpha n}E_{f}f_{\alpha n}^{+}f_{\alpha n}
\nonumber \\
&&+\sum_{\alpha n}(Vb_{n}^{+}c_{\alpha n}^{+}f_{\alpha n}+H.c.),
\label{Anderson}
\end{eqnarray}
where $c_{\alpha {\bf k}}^{+}$ and $c_{\alpha {\bf k}}$ are creation and
annihilation operators of conduction band holes with $z$-component $\alpha $
of the total magnetic moment, $\alpha =-j,-j+1,...j$, and wave vector {\bf %
k. } Operators $f_{\alpha n}^{+}$ and $f_{\alpha n}$ create and annihilate $%
f $ holes on a lattice site with a number $n$. The Bose operators $b_{n}^{+}$
and $b_{n}$ are the conventional slave bosons. A wave function of the state
with one boson on a site $n$ is equal to $b_{n}^{+}\mid O>.$ This state
corresponds to the 4$f^{14}$ configuration of the Yb$^{+2}$ ion on the site $%
n$. The operator $b_{n}^{+}b_{n}$ determines the probability to find the $f$
ion in the divalent state. In order to admit valent fluctuations only
between Yb$^{+3}$ and Yb$^{+2}$ states the following constraint is imposed
on each lattice site:\
\begin{equation}
b_{n}^{+}b_{n}+\sum_{\alpha }f_{\alpha n}^{+}f_{\alpha n}=1.
\label{constrain}
\end{equation}
We should mention that in the limit $\mu -E_{f}<<\pi \left| V\right|
^{2}\rho _{0}$ ($\mu $ is the chemical potential and $\rho _{0}$ is a
conduction band density of states) one can integrate over the bosons. As a
result, this model becomes the $SU(N)$ periodic Coqblin-Schriffer model with
an exchange interaction $J=\left| V\right| ^{2}/(\mu -E_{f})$:\cite
{NewsRead,Bickers87}
\begin{equation}
H_{A}=\sum_{a{\bf k}}\varepsilon _{{\bf k}}c_{\alpha {\bf k}}^{+}c_{\alpha
{\bf k}}-J\sum_{\alpha \beta n}(c_{\alpha n}^{+}f_{\alpha n})(f_{\beta
n}^{+}c_{\beta n})  \label{c-s-model}
\end{equation}
This model describes the integral valence case when the $f^{13}$ state
behaves like a spin interacting antiferromagnetically with conduction band
holes$.$ In mixed-valence compounds $E_{f}$ can be close to $\mu $, i.e. $%
\mu -E_{f}\sim \pi \left| V\right| ^{2}\rho _{0}$.\cite{Schlottmann}
Therefore, in the latter case it is necessary to use the lattice Anderson
model.

When a $f$ hole transits into the conduction band, then a corresponding Yb
ion acquires an excess negative charge ($\delta q=e$). The excess charge
produces a Coulomb potential that attracts conduction band holes. Due to the
small charge carrier concentration ($N_{0c}=0.07$ per formula unit) the
interaction is not completely screened. At least, the Coulomb interaction
between the fluctuating charge of Yb ions and electrons (or holes) on
neighboring Cu ions of which $p$ and $d$ states mainly contribute to the
formation of the conduction band, should be taken into account.
Unfortunately, a detailed consideration of the long range Coulomb
interaction is a complicated problem. For the sake of simplicity we
approximate the interaction by use of an on-site Coulomb attraction between
the excess charge and conduction holes. The same approximation is made
within the promotional models.\cite{F-K,R-F} In the slave boson
representation this on-site attraction can be written as follows:\
\begin{equation}
H_{C}=-U_{cf}\sum_{\alpha n}b_{n}^{+}b_{n}c_{\alpha n}^{+}c_{\alpha n},
\label{attraction}
\end{equation}
where the parameter $U_{cf}>0$. It characterizes the effective value of the
Coulomb interaction between 4$f$ and conduction band holes. Using the
constraint Eq.(\ref{constrain}), one can rewrite the interaction in the
form\
\begin{equation}
H_{C}=U_{cf}\sum_{\alpha \beta n}f_{\beta n}^{+}f_{\beta n}c_{\alpha
n}^{+}c_{\alpha n}-U_{cf}\sum_{\alpha n}c_{\alpha n}^{+}c_{\alpha n}
\label{repulsion}
\end{equation}
The first term describes an on-site Coulomb repulsion between conduction
band and $f$ holes. The second term results in an energy shift of the
conduction band. According to the representation (\ref{repulsion}), the
attractive interaction Eq.(\ref{attraction}) is equivalent to the on-site
repulsion between conduction band and $f$ electrons. In the present paper we
prefer to use the representation Eq.(\ref{attraction}).

The total Hamiltonian of the system under consideration is given by
\begin{equation}
H=H_A+H_C.  \label{Hamiltonian}
\end{equation}
The model described by the Hamiltonian can be solved exactly in the limit of
large degeneracy $N=2j+1>>1$ if we suppose the following dependence of the
parameters $V$ and $U_{cf}$ on $N$:
\begin{equation}
V=\widetilde{V}N^{-1/2},\text{ }U_{cf}=\widetilde{U}_{cf}N^{-1},
\label{scaling}
\end{equation}
where $\widetilde{V}$ and $\widetilde{U}_{cf}$ are finite in the limit $N>>1$%
. Moreover, it has to be supposed that the initial number of conduction band
holes, $n_{0c}=N_{0c}/N$, and the number of $f$ holes, $n_{0f}=N_{0f}/N$,
per unit cell and orbital are finite in the limit $N>>1$. The constrain Eq.(%
\ref{constrain}) must be replaced by the following one:\cite{Coleman87}
\begin{equation}
b_n^{+}b_n+\sum_\alpha f_{\alpha n}^{+}f_{\alpha n}=Nn_{0f}\text{ ,}
\label{constrain2}
\end{equation}
where $n_{0f}=1/8$ for the considered Yb system.

The partition function of the model Eq.(\ref{Hamiltonian}) may be written as
a path integral over Grassmann variables $c^{+}$, $c$, $f$ $^{+}$,$f$ and
Bose variables $b^{+}$, $b$ and $\lambda :\;$%
\begin{equation}
Z=\int D(c^{+}cf^{+}fb^{+}b\lambda )\exp (-\int\limits_{0}^{\beta }d\tau
L(\tau )).  \label{Z}
\end{equation}
Here the Lagrangian $L$ is given by\
\begin{eqnarray}
L(\tau ) &=&\sum_{a{\bf k}}c_{\alpha {\bf k}}^{+}(\partial _{\tau
}+\varepsilon _{{\bf k}}-\mu )c_{\alpha {\bf k}}+\sum_{n}b_{n}^{+}\partial
_{\tau }b_{n}  \nonumber \\
&&+\sum_{\alpha n}\{f_{\alpha n}^{+}(\partial _{\tau }+E_{f}-\mu )f_{\alpha
n}+(Vb_{n}^{+}c_{\alpha n}^{+}f_{\alpha n}  \nonumber \\
&&+H.c.)-U_{cf}b_{n}^{+}b_{n}c_{\alpha n}^{+}c_{\alpha n}\}+\sum_{n}i\lambda
_{n}(b_{n}^{+}b_{n}  \nonumber \\
&&+\sum_{\alpha }f_{\alpha n}^{+}f_{\alpha n}-Nn_{0f}),  \label{Lagr}
\end{eqnarray}
where all variables $c_{\alpha n}^{+}$, $c_{\alpha n}$, $f_{\alpha n}^{+}$,$%
f_{\alpha n}$ , $b_{n}^{+}$, $b_{n}$ and $\lambda _{n}$ are functions of the
imaginary time $\tau .$ Let us use the following transformation :\
\[
\exp (U_{cf}\sum_{\alpha n}\int\limits_{0}^{\beta }b_{n}^{+}b_{n}c_{\alpha
n}^{+}c_{\alpha n}d\tau )=
\]
\begin{eqnarray}
&&U_{cf}\pi ^{-1}\int D\Psi _{n}D\Psi _{n}^{*}\exp
(-U_{cf}\sum_{n}\int\limits_{0}^{\beta }(\Psi _{n}\Psi _{n}^{*}  \nonumber \\
&&+\Psi _{n}b_{n}^{+}b_{n}+\Psi _{n}^{*}\sum_{\alpha }c_{\alpha
n}^{+}c_{\alpha n})d\tau ,  \label{trans}
\end{eqnarray}
where $\Psi _{n}(\tau )$ and $\Psi _{n}^{*}(\tau )$ are complex conjugate
Bose variables. Then, substituting this transformation into Eq.(\ref{Lagr}),
we obtain the partition function as a path integral over the Grassmann
variables $c_{\alpha n}^{+}$, $c_{\alpha n}$, $f_{\alpha n}^{+}$,$f_{\alpha
n}$ and Bose variables $b_{n}^{+}$, $b_{n}$, $\Psi $, $\Psi ^{*}$ and $%
\lambda _{n}$. The integration over the Bose variables can be performed
using the saddle-point method that gives an exact solution of the model with
the Hamiltonian (\ref{Hamiltonian}) in the limit $N\rightarrow \infty $. We
look for an uniform saddle-point solution:
\begin{eqnarray}
i\lambda _{n}(\tau ) &=&\lambda ,\text{ \ }b_{n}(\tau )=b,  \nonumber \\
\Psi _{n}(\tau ) &=&\Psi ,\text{ }\Psi _{n}^{*}(\tau )=\widetilde{\Psi }.
\label{s-point}
\end{eqnarray}
Here it should be noted that in the saddle point the $\widetilde{\Psi }$ is
not complex conjugate to $\Psi $, i.e$.$ $\Psi ^{*}\neq $ $\widetilde{\Psi }$%
, because Re$\Psi $ and Im$\Psi $ must be considered as independent
variables. The physical meaning of $b$ , $\Psi $ and $\widetilde{\Psi }$
becomes clear from the mean field equations:
\begin{equation}
b=-\frac{V}{\left( \lambda +U_{cf}\Psi \right) N_{u}}\sum_{\alpha {\bf k}%
}\left\langle c_{\alpha {\bf k}}^{+}f_{\alpha {\bf k}}\right\rangle _{MF},
\label{hybridization}
\end{equation}
\begin{equation}
\Psi =-\sum_{\alpha }\left\langle c_{\alpha n}^{+}c_{\alpha n}\right\rangle
_{MF}=-N_{c},  \label{psi}
\end{equation}
\begin{equation}
\widetilde{\Psi }=-\left| b\right| ^{2},  \label{psi-c}
\end{equation}
where $N_{u}$ is the number of unit cells in the lattice, $N_{c}$ is the
number of conduction band holes per unit cell , $<...>_{MF}$ means averaging
over the mean-field Hamiltonian $H_{MF}$ that easily follows from Eqs.(\ref
{Lagr})-(\ref{psi-c}):
\begin{eqnarray}
H_{MF} &=&\sum_{a{\bf k}}\{(\varepsilon _{{\bf k}}-U_{cf}\left| b\right|
^{2})c_{\alpha {\bf k}}^{+}c_{\alpha {\bf k}}+\varepsilon _{f}f_{\alpha {\bf %
k}}^{+}f_{\alpha {\bf k}}  \nonumber \\
&&+(Vb^{*}c_{\alpha {\bf k}}^{+}f_{\alpha {\bf k}}+H.c.)\}-N_{u}(\varepsilon
_{f}-E_{f})N_{f},  \label{MFH2}
\end{eqnarray}
where $\varepsilon _{f}$ is the renormalized $f$ level energy:
\begin{equation}
\varepsilon _{f}\equiv E_{f}+\lambda \text{ ,}  \label{f-energy}
\end{equation}
$N_{f}\equiv \sum_{\alpha }\left\langle f_{\alpha n}^{+}f_{\alpha
n}\right\rangle $ is the average number of $f$ holes on the $f$ shell on the
site $n$. The parameter determines the valence ($\nu )$ of the $f$ ion. For
example, for an Yb ion we have $\nu =2+N_{f}.$ Below, for the sake of
simplicity we shall assume that there is only one Yb ion per unit cell. The
order parameter $b$ describes the formation of the HF state (in other words,
the coherent Kondo state). According to the constraint Eq.(\ref{constrain})
(or Eq.(\ref{constrain2})) there is the following relation between $b$ and a
valence change ($\Delta N_{f}$) of $f$ ions:
\begin{equation}
\left| b\right| ^{2}=Nn_{0f}-\sum_{\alpha }\left\langle f_{\alpha
n}^{+}f_{\alpha n}\right\rangle =N_{0f}-N_{f}\equiv \Delta N_{f}
\label{b-valence}
\end{equation}
Because in the HF state $\left| b\right| \neq 0,$ we come to the very
important conclusion that in the HF state $f$ ions have a non-integral
valence. For example, for Yb ions we have $N_{0f}=1$, and the valence is
equal to $\nu =3-\Delta N_{f}.$ Strictly speaking, due to valence
fluctuations governed by the hybridization terms in the Anderson Hamiltonian
Eq.(\ref{Anderson}), even in the high temperature region where $\left\langle
b_{n}\right\rangle =0$ the valency is not integral, i.e. $%
N_{f}=N_{0f}-\left\langle b_{n}^{+}b_{n}\right\rangle <N_{0f},$ because $%
\left\langle b_{n}^{+}b_{n}\right\rangle \neq 0.$ In order to take into
account the effect it is necessary to go beyond the mean-field approach and
calculate $1/N$ corrections.

It is convenient instead of the dimensionless order parameter $b$ to use the
following energy parameter characterizing the effective hybridization:
\begin{equation}
B\equiv V^{*}b\text{ .}  \label{B-definition}
\end{equation}
Inserting Eq.(\ref{B-definition}) into Eq.(\ref{b-valence}) gives a useful
relation:\
\begin{equation}
\left| B\right| ^{2}=\left| V\right| ^{2}\Delta N_{f}\text{ .}  \label{B2}
\end{equation}
Then Eq.(\ref{hybridization}) takes the conventional form:
\begin{equation}
B=-\frac{J}{N_{u}}\sum_{\alpha {\bf k}}\left\langle c_{\alpha {\bf k}%
}^{+}f_{\alpha {\bf k}}\right\rangle \text{ ,}  \label{hybrid2}
\end{equation}
where the exchange interaction $J$ is given by:
\begin{equation}
J=\frac{\left| V\right| ^{2}}{\varepsilon _{f}-E_{f}-U_{cf}N_{c}}.
\label{coupling}
\end{equation}
From Eq.(\ref{MFH2}) it follows that the interaction $U_{cf}$ leads to the
energy shift $U_{cf}\Delta N_{f\text{ }}$ of the conduction band to lower
energies. This in turn shifts the chemical potential $\mu $ and effective
energy $\varepsilon _{f}$ with respect to their bare values $\mu _{0}$ and $%
\varepsilon _{0f}$.

In accordance to Eq.(\ref{coupling}) the interaction $U_{cf}$ influences the
exchange interaction $J$, which in turn effects on the Kondo screening. This
effect plays a very important role and will be discussed below in detail.
From the physical point of view the effect is related to the renormalization
of the energy of the first excited state of Yb ions due to the Coulomb
interaction.

The diagonalization of the Hamiltonian $H_{MF}$ shows that in the HF state
the renormalized band structure consists of two hybrid bands:\
\begin{equation}
\widetilde{E}_{\eta {\bf k}}=\frac{1}{2}\{\varepsilon _{{\bf k}}+\widetilde{%
\varepsilon }_{f}\mp [(\varepsilon _{{\bf k}}-\widetilde{\varepsilon }%
_{f})^{2}+4\left| B\right| ^{2}]^{1/2}\}.  \label{h-bands0}
\end{equation}
where the upper and lower signs corresponds to $\eta =1$ and $2$,
respectively, and the following energy parameters are introduced:\
\begin{eqnarray}
\widetilde{\varepsilon }_{f} &\equiv &\varepsilon _{f}+U_{cf}\Delta N_{f},
\nonumber \\
\widetilde{E}_{\eta {\bf k}} &=&E_{\eta {\bf k}}+U_{cf}\Delta N_{f},
\label{e-shifted} \\
\widetilde{\mu } &\equiv &\mu +U_{cf}\Delta N_{f}.  \nonumber
\end{eqnarray}
The free energy of the system is determined by the well known equation:\
\begin{equation}
F=E-ST  \label{FreeEnergy}
\end{equation}
The internal energy $E$ and entropy $S$ of the electron system per unit cell
(or per $f$ ion, because we assume that there is only one $f$ ion per unit
cell) is equal to\
\begin{eqnarray}
E &=&\frac{N}{N_{u}}\sum_{\eta =1,2}\sum_{{\bf k}}\widetilde{E}_{\eta {\bf k}%
}f(\widetilde{E}_{\eta {\bf k}})-U_{cf}\Delta N_{f}N_{t}  \label{energy0} \\
&&-(\varepsilon _{f}-E_{f})N_{f},  \nonumber
\end{eqnarray}

\begin{eqnarray}
S &=&-\frac{N}{N_{u}}\sum_{\eta =1,2}\sum_{{\bf k}}\{f(\widetilde{E}_{\eta
{\bf k}})\ln f(\widetilde{E}_{\eta {\bf k}})  \label{entropy0} \\
&&+(1-f(\widetilde{E}_{\eta {\bf k}}))\ln (1-f(\widetilde{E}_{\eta {\bf k}%
}))\},  \nonumber
\end{eqnarray}
where
\begin{equation}
N_{t}=N_{c}+N_{f}=N_{0c}+N_{0f}  \label{total}
\end{equation}
is the total number of holes per unit cell, and $f(\widetilde{E})=(\exp ((%
\widetilde{E}-\widetilde{\mu })/T)+1)^{-1}=f(E)$. It is interesting to note
that $N_{c}=N_{0c}+\Delta N_{f},$ which demonstrates the fact that a certain
part of holes from $f$ shells transit into the conduction band.

\section{Mean-field equations}

The state of the system at any temperature $T$ is completely determined by
three parameters: $\widetilde{\mu }$, $B$ and $\widetilde{\varepsilon }_{f}$%
. To determine the parameters it is necessary to solve a set of three
nonlinear mean-field equations:\
\begin{equation}
N_{t}=\frac{1}{N_{u}}\sum_{\alpha {\bf k}}\left( \left\langle c_{\alpha {\bf %
k}}^{+}c_{\alpha {\bf k}}\right\rangle +\left\langle f_{\alpha {\bf k}%
}^{+}f_{\alpha {\bf k}}\right\rangle \right) ,  \label{eq-1}
\end{equation}
\begin{equation}
N_{f}=\frac{1}{N_{u}}\sum_{\alpha {\bf k}}\left\langle f_{\alpha {\bf k}%
}^{+}f_{\alpha {\bf k}}\right\rangle ,  \label{eq-2}
\end{equation}
\begin{equation}
B=-\frac{J}{N_{u}}\sum_{\alpha {\bf k}}\left\langle c_{\alpha {\bf k}%
}^{+}f_{\alpha {\bf k}}\right\rangle .  \label{eq-3}
\end{equation}
At arbitrary $T$ \ the mean-field equations have a trivial solution $B=0$.
Moreover, nontrivial solutions with $B\neq 0$ can also exists. These
solutions correspond to either minimum or maximum of the free energy $F$. In
the case $B\neq 0$ one can rewrite the mean-field equations in a form:\
\begin{equation}
N_{t}=N\int d\varepsilon \rho (\varepsilon )(f(\widetilde{E}_{1{\bf k}})+f(%
\widetilde{E}_{2{\bf k}})),  \label{eq1n}
\end{equation}
\begin{equation}
N_{f}=N\int d\varepsilon \rho (\varepsilon )\frac{\left| B\right| ^{2}f(%
\widetilde{E}_{1{\bf k}})+(\widetilde{\varepsilon }_{f}-\widetilde{E}_{1{\bf %
k}})^{2}f(\widetilde{E}_{2{\bf k}})}{\left| B\right| ^{2}+(\widetilde{%
\varepsilon }_{f}-\widetilde{E}_{1{\bf k}})^{2}},  \label{eq2n}
\end{equation}
\begin{equation}
\frac{1}{J}=N\int d\varepsilon \rho (\varepsilon )\frac{f(\widetilde{E}_{1%
{\bf k}})-f(\widetilde{E}_{2{\bf k}})}{\widetilde{E}_{2{\bf k}}-\widetilde{E}%
_{1{\bf k}}},  \label{eq3n}
\end{equation}
where $\rho (\varepsilon )$ is the density of states (DOS) in the conduction
band:\
\begin{equation}
\rho (\varepsilon )=\frac{1}{N_{u}}\sum_{{\bf k}}\delta (\varepsilon
-\varepsilon _{{\bf k}}).  \label{DOS}
\end{equation}
It is convenient to represent the internal energy $E$, Eq.(\ref{energy0}),
as follows:\
\begin{eqnarray}
E &=&N\sum_{\eta =1,2}\int d\varepsilon \rho (\varepsilon )\widetilde{E}%
_{\eta {\bf k}}f(\widetilde{E}_{\eta {\bf k}})+\frac{1}{J}\left| B\right|
^{2}  \nonumber \\
&&+\frac{U_{cf}}{\left| V\right| ^{4}}\left| B\right| ^{4}-(\widetilde{%
\varepsilon }_{f}-E_{f})N_{0f}.  \label{e4-internal}
\end{eqnarray}
Let us analyze the exchange interaction $J$ given by Eq.(\ref{coupling}).
Using the relations $N_{c}=N_{0c}+\Delta N_{f}$ and Eq.(\ref{B2}), one can
rewrite $J^{-1}$ in the form:\
\begin{equation}
\frac{1}{J}=\frac{1}{\left| V\right| ^{2}}\left( \widetilde{\varepsilon }%
_{f}-E_{f}-U_{cf}N_{0c}-\frac{2U_{cf}B^{2}}{\left| V\right| ^{2}}\right)
\label{coupling-g}
\end{equation}
Eq.(\ref{coupling-g}) shows that in the HF state, i.e. at $B\neq 0,$ the
exchange interaction $J$ is enhanced with respect to one in the normal state
with $B=0.$ It is the effect that is the driving force of the first-order
valence phase transition in YbInCu$_{4}$ as it will be shown below.

In the case $U_{cf}=0$ at $T<T_{k}$ the mean-field equations Eqs.(\ref{eq1n}%
)-(\ref{eq3n}) have only one nontrivial solution with $B\neq 0$ which
describes a continuous transition into the HF state. We call the solution
``conventional''. The Kondo temperature $T_{k}$ can be found from Eq.(\ref
{eq3n}), supposing $B=0.$ A detailed analysis of the HF state described by
the conventional solution can be found, for example, in the review.\cite
{NewsRead} Temperature behavior of $\Delta N_{f}$ and susceptibility in the
HF state found from a numerical solution of Eqs.(\ref{eq1n})-(\ref{eq3n})
will be discussed below in Sec. VI.

One can consider the formation of the Fermi-liquid state with a non-zero
effective hybridization $B$ as  a coherent Kondo effect .\cite{NewsRead} The
mean-field theory enables us to find the leading contribution in terms the $%
1/N$ expansion into the phenomena. In the high temperature region the
mean-field solution with $B=0$ discribes a localized $f$ holes that do not
interact with conduction holes. To take into account the interaction it is
necessary to consider $1/N$ corrections due to slave boson fluctuations
around the mean-field solution. The corrections can be important in the
region of crossover from a weak coupling regime to a strong coupling regime
that occurs at $T\sim T_{k}.$ \cite{ColemanAndrei}

In addition, one can note that at $T=T_k$ the Green's function $%
<b_n^{+}(i\omega )b_n(i\omega )>$ has a pole at $\omega =0$. It is the
singularity that results in the Abrikosov-Suhl resonance.

At low temperatures $T<<T_{k}$ in the HF state a new energy scale arises. It
is the so called low temperature Kondo scale $T_{0}$ determined as\
\begin{equation}
T_{0}\equiv \widetilde{\varepsilon }_{f}-\widetilde{\mu }.  \label{T0}
\end{equation}
Solving Eq.(\ref{eq3n}) at $T=0$ in the case $\rho (\varepsilon )=\rho _{0}$
gives\
\begin{equation}
T_{0}=(\widetilde{\varepsilon }_{f}-\widetilde{E}_{\text{min}})\exp (-\frac{1%
}{N\rho _{0}J})  \label{T01}
\end{equation}
where $\widetilde{E}_{\text{min}}=\widetilde{E}_{\text{1}}(k=0).$ It is
important to note a difference between the two Kondo scales $T_{0}$ and $%
T_{k}$. But the Abrikosov-Suhl resonance, the Kondo temperature $T_{k}$
appears also in a high-temperature expansion of the spin susceptibility, for
example, and other physical parameters in the weak coupling regime. $T_{0}$
is an energy scale that determines thermodynamic properties in the
Fermi-liquid regime with strong coupling. In general case, $T_{0}\neq T_{k}$
(for a detail discussion on the problem see, for example, a review \cite
{Bickers87}).

\section{Striction phenomena}

When an Yb ion transits from the ground state Yb$^{+3}$ into the excited
state Yb$^{+2},$ the radius of the $f$ shell increases. It results in an
local internal pressure that brings about a local lattice strain. Let $%
e_{ij} $, where $i,j=x,y,z,$ be a local strain tensor. Since the probability
to find an Yb ion on the site $n$ in the divalent state is given by the
operator $b_{n}^{+}b_{n}$, we can write the energy related to this type of
the electron-lattice interaction in the form\
\begin{equation}
H_{e-l}=-\sum_{n}d_{ij}e_{ji}(n)b_{n}^{+}b_{n}.  \label{e-l-energy}
\end{equation}
For a cubic lattice the energy tensor $d_{ij}$ has a simple form $%
d_{ij}=\delta _{ij}d$ , therefore this electron-lattice interaction can be
rewritten as:\
\begin{equation}
H_{e-l}=-d\sum_{n}e_{B}(n)b_{n}^{+}b_{n}.  \label{e-l-energy2}
\end{equation}
where $e_{B}(n)=e_{xx}(n)+e_{yy}(n)+e_{zz}(n)$ is a local volume strain. For
Yb compounds the interaction energy $d$ is positive, which corresponds to a
local volume expansion when Yb ions change their valence from +3 to +2. In
turn, $d$ is negative for Ce ions with the ground state configuration $%
4f^{1},$ because the $f$ shell of a Ce ion in the excited $4f^{0}$ state has
a smaller radius than in the $4f^{1}$ state. In order to make clear a
physical meaning of the interaction $H_{e-l}$ let us find an average value $%
<H_{e-l}>$ in the case of an uniform strain $e_{B}(n)=e_{B}.$ Taking into
account Eq.(\ref{b-valence}) we find\
\begin{equation}
<H_{e-l}>=-de_{B}\Delta N_{f}N_{u}  \label{el-energy3}
\end{equation}
The volume strain $e_{B}$ is related to the volume change ($\Delta v)$ of
the whole volume $(v),$ namely, $e_{B}=\Delta v/v$. Then Eq.(\ref{el-energy3}%
) takes the form\
\begin{equation}
<H_{e-l}>=p_{0}\Delta N_{f}\Delta v  \label{el-4}
\end{equation}
where we introduce a new fundamental parameter\
\begin{equation}
p_{0}\equiv -d/v_{0}.  \label{p0}
\end{equation}
Here $v_{0}$ is the unit cell volume (per $f$ ion). The parameter $p_{0}$
has the following physical meaning: $p_{0}$ is a pressure produced by a $f$
ion on the lattice when the $f$ ion valence is changed by 1. $p_{0}$ is
negative for Yb ions, because it tends to expand the lattice, and positive
for Ce ions, because it tends to shrink the lattice. The energy Eq.(\ref
{el-4}) has a simple physical meaning. This is the work produced by the
pressure $p_{0}\Delta N_{f}$ in order to change the volume of the system by
value $\Delta v$.

Apart of the $f$-shell-size-fluctuation mechanism (Eq.(\ref{e-l-energy})) of
the electron-lattice interaction there is another mechanism related to a
volume dependence of the hybridization parameter $V$ in Eq.(\ref{Anderson}):

\begin{equation}
V(e_{B})=V\exp (re_{B})  \label{hybrid}
\end{equation}
where $r<0$ that corresponds to increasing $V(e_{B})$ when decreasing
volume. Eq.(\ref{hybrid}) follows from a volume dependence of overlapping
between $f$ and conduction band wave functions. Usually it is supposed that
it is the mechanism that is responsible for a pressure dependence of $T_{k}$
and the Kondo volume collapse phenomena in compounds with Kondo impurities
and HF compounds. However, as we shall show below it is not valid in the
case of Yb compounds in which the $f$-shell-size-fluctuation mechanism plays
a more important role. Taking into account these two mechanisms of the
electron-lattice interaction represented by Eqs.(\ref{e-l-energy2}) and (\ref
{hybrid}), we obtain the following total Hamiltonian:\
\begin{equation}
H=H_{A}+H_{C}+H_{e-l}+H_{lat},  \label{t-Hamiltonian}
\end{equation}
where\
\begin{equation}
H_{lat}=\frac{1}{2}C_{B}e_{B}^{2}N_{u}  \label{lat-energy}
\end{equation}
is the energy of a uniformly deformed lattice. $C_{B}$ is the bulk modulus
per $f$ ion. The Hamiltonian Eq.(\ref{t-Hamiltonian}) results in the same
mean-field equations (\ref{eq1n})-(\ref{eq3n}) with the only difference.
Namely, the exchange interaction $J$ becomes strain dependent:\
\begin{equation}
J(e_{B})=\frac{\left| V(e_{B})\right| ^{2}}{\widetilde{\varepsilon }%
_{f}-E_{f}-U_{cf}N_{0c}-2U_{cf}\Delta N_{f}-de_{B}}  \label{coupling3a}
\end{equation}
Taking into account the elastic energy Eq.(\ref{lat-energy}) and electron
energy Eq.(\ref{e4-internal}), the total internal energy per $f$ ion can be
written in the form\
\begin{equation}
E_{t}=E+\frac{1}{2}C_{B}e_{B}^{2}.  \label{et-internal}
\end{equation}
In the equilibrium state the strain $e_{B}$ is determined by a minimization
of the free energy with respect to $e_{B}:$ $\partial F/\partial e_{B}=0.$
Taking into account Eqs.(\ref{coupling3a}) and (\ref{hybrid}) the
minimization results in an equation:\
\begin{equation}
e_{B}=(\frac{d}{C_{B}}+\frac{2r\left| V(e_{B})\right| ^{2}}{C_{B}J(e_{B})}%
)\Delta N_{f}  \label{strain-g}
\end{equation}
which easily follows from the relation:\
\begin{eqnarray*}
\partial F/\partial e_{B} &=&\partial E_{t}/\partial e_{B} \\
&=&\left| B\right| ^{2}\frac{\partial }{\partial e_{B}}(\frac{1}{J(e_{B})}%
)+U_{cf}\left| B\right| ^{4}\frac{\partial }{\partial e_{B}}(\frac{1}{\left|
V(e_{B})\right| ^{4}}) \\
&&+C_{B}e_{B}
\end{eqnarray*}
since the entropy $S$, Eq.(\ref{entropy0}), does not depend on $e_{B}$ in a
direct way (here we neglect a volume dependence of the Fermi energy). Eq.(%
\ref{strain-g}) shows that the volume strain $e_{B}$ is proportional to the
valence change $\Delta N_{f}$ of the Yb ions. Such a proportionality have
been observed experimentally.\cite{Cornelius97} Eq.(\ref{strain-g}) together
with Eqs.(\ref{eq1n})-(\ref{eq3n}), where $J$ is replaced by $J(e_{B}),$
represent a closed set of nonlinear equations which determine the
equilibrium parameters $\widetilde{\varepsilon }_{f}(T)$, $\widetilde{\mu }%
(T),$ $B(T)$ and $e_{B}(T).$

In the case of a sufficiently small strain $\left| e_{B}\right| <<1$ one can
use a linear approximation:
\begin{equation}
J(e_{B})^{-1}\approx J(0)^{-1}+N\rho _{0}\Omega e_{B},  \label{coupl3}
\end{equation}
where $\rho _{0}$ is the DOS on the Fermi surface, $\Omega =\Omega
_{1}+\Omega _{2}+\Omega _{3}$,
\begin{equation}
\Omega _{1}+\Omega _{2}=%
%TCIMACRO{\QOVERD( ) {\partial \ln J(e_{B})}{\partial \ln v} }
%BeginExpansion
{\partial \ln J(e_{B}) \overwithdelims() \partial \ln v}%
%EndExpansion
_{B,\widetilde{\varepsilon }_{f},e_{B}=0}  \label{Grueneisen}
\end{equation}

\[
\Omega _{1}=-\frac{2r}{N\rho _{0}J},\text{ }\Omega _{2}=-\frac{d}{N\rho
_{0}\left| V\right| ^{2}}.
\]
\[
\Omega _{3}=\frac{1}{N\rho _{0}}\left( \frac{\partial J^{-1}}{\partial B}%
\frac{\partial B}{\partial e_{B}}+\frac{\partial J^{-1}}{\partial \widetilde{%
\varepsilon }_{f}}\frac{\partial \widetilde{\varepsilon }_{f}}{\partial e_{B}%
}\right) _{e_{B}=0}
\]
Since $r<0,$ consequently, $\Omega _{1}$ is always positive while $\Omega
_{2}$ can be both positive and negative. In general case $\Omega $ can be
both positive and negative. The physical meaning of the parameter $\Omega $
comes from Eq.(\ref{T01}). Substitution of Eq.(\ref{coupl3}) into Eq.(\ref
{T01}) gives\
\begin{equation}
T_{0}(e_{B})\sim \exp (-\frac{1}{N\rho _{0}J(0)}-\Omega e_{B}).
\label{Kondo0s}
\end{equation}
The equation determines the dependence of the low temperature Kondo scale $%
T_{0}$ on the strain $e_{B}$.

Electron-lattice interactions described by Eqs.(\ref{e-l-energy}) and (\ref
{hybrid}) lead to a renormalization of the elastic constants. In order to
find the effect it is necessary to calculate the free energy $F$ as a
function of $e_{B}$ at given $T$ : $F=F(T,e_{B}).$ For this purpose we have
to solve Eqs.(\ref{eq1n})-(\ref{eq3n}) with the coupling Eq.(\ref{coupling3a}%
) at fixed $e_{B}.$ This enables us to find $\widetilde{\varepsilon }%
_{f}(T,e_{B})$, $\widetilde{\mu }(T,e_{B})$ and $B(T,e_{B})$ and then $%
F(T,e_{B})$. Renormalized elastic constants $c_{ij}^{*}$ are equal to\
\begin{equation}
c_{ij}^{*}=\left( \frac{\partial ^{2}F}{\partial e_{ii}\partial e_{jj}}%
\right) _{e=e(T)},  \label{el-const}
\end{equation}
where $e_{ii}(T)=e_{B}(T)/3$ is the equilibrium elastic strain.

An effect of applied pressure $P$ on the system under consideration can be
calculated from the thermodynamic relation\
\begin{equation}
P=-\frac{\partial F(v,T)}{\partial v}  \label{pressure}
\end{equation}
Eq.(\ref{pressure}) together with Eqs.(\ref{eq1n})-(\ref{eq3n}), where $J$
is replaced by $J(e_{B}),$ represent a closed set of nonlinear equations
that determine the equilibrium parameters $\widetilde{\varepsilon }_{f}(T,P)$%
, $\widetilde{\mu }(T,P),$ $B(T,P)$ and $e_{B}(T,P).$

\section{Magnetic field effect}

Let us study an influence of a magnetic field $H$ on the system and the
valence phase transition. We shall take into account only an interaction of
spins of conduction band and localized $f$ holes with the magnetic field
(Zeeman energy). The Zeeman energy of the spins in a magnetic field $H$
directed along $z$ axis is equal to\
\begin{equation}
H_{z}=-g\mu _{B}H\sum_{n}(S_{fn}^{z}+S_{cn}^{z}),  \label{Zeeman}
\end{equation}
where $g$ and $\mu _{B}$ are the gyromagnetic factor and the Bohr magneton,
respectively, $S_{cn}^{z}$ and $S_{fn}^{z}$ are $z$ components of the spin
operators for conduction band and $f$ holes: \
\begin{eqnarray}
S_{cn}^{z} &=&\sum_{\alpha =-j}^{j}\alpha c_{\alpha n}^{+}c_{\alpha n},
\nonumber \\
S_{fn}^{z} &=&\sum_{\alpha =-j}^{j}\alpha f_{\alpha n}^{+}f_{\alpha n}.
\label{z-spin}
\end{eqnarray}
The magnetic field splits the $N$-fold degenerate hybrid bands Eq.(\ref
{h-bands0}):\
\begin{equation}
E_{\eta \alpha {\bf k}}=E_{\eta {\bf k}}-\alpha g\mu _{B}H.  \label{h-bands}
\end{equation}
Neglecting a magnetic field effect on the orbital motion of heavy fermions,
which is possible due to the large mass of the quasiparticles, one can write
the mean-field equations (\ref{eq1n})-(\ref{eq3n}) in the form:\
\begin{equation}
N_{t}=\sum_{\alpha }\int d\varepsilon \rho (\varepsilon )(f(\widetilde{E}%
_{1\alpha {\bf k}})+f(\widetilde{E}_{2\alpha {\bf k}})),  \label{eq1h}
\end{equation}
\begin{equation}
N_{f}=\sum_{\alpha }\int d\varepsilon \rho (\varepsilon )\frac{\left|
B\right| ^{2}f(\widetilde{E}_{1\alpha {\bf k}})+(\widetilde{\varepsilon }%
_{f}-\widetilde{E}_{1{\bf k}})^{2}f(\widetilde{E}_{2\alpha {\bf k}})}{\left|
B\right| ^{2}+(\widetilde{\varepsilon }_{f}-\widetilde{E}_{1{\bf k}})^{2}},
\label{eq2h}
\end{equation}
\begin{equation}
\frac{1}{J}=\sum_{\alpha }\int d\varepsilon \rho (\varepsilon )\frac{f(%
\widetilde{E}_{1\alpha {\bf k}})-f(\widetilde{E}_{2\alpha {\bf k}})}{%
\widetilde{E}_{2{\bf k}}-\widetilde{E}_{1{\bf k}}}.  \label{eq3h}
\end{equation}
If we take into account the electron-lattice interaction then the exchange
interaction $J$ in Eq.(\ref{eq3h}) is given by Eq.(\ref{coupling3a}).
Solving these equations together with Eq.(\ref{strain-g}), we can find
parameters $\widetilde{\varepsilon }_{f}(T,H)$, $\widetilde{\mu }(T,H),$ $%
B(T,H)$ and $e_{B}(T,H)$. Then we can calculate the total internal energy
per $f$ ion:\
\begin{eqnarray}
E_{t} &=&\sum_{\alpha }\sum_{\eta =1,2}\int d\varepsilon \rho (\varepsilon )%
\widetilde{E}_{\eta \alpha {\bf k}}f(\widetilde{E}_{\eta \alpha {\bf k}})+%
\frac{1}{J(e_{B})}\left| B\right| ^{2}  \nonumber \\
&&+\frac{U_{cf}}{\left| V(e_{B})\right| ^{4}}\left| B\right| ^{4}-(%
\widetilde{\varepsilon }_{f}-E_{f})N_{0f}+\frac{1}{2}C_{B}e_{B}^{2},
\label{et-h-internal}
\end{eqnarray}
and the entropy
\begin{eqnarray}
S &=&-\sum_{\alpha }\sum_{\eta =1,2}\int d\varepsilon \rho (\varepsilon )\{f(%
\widetilde{E}_{\eta \alpha {\bf k}})\ln f(\widetilde{E}_{\eta \alpha {\bf k}%
})  \nonumber \\
&&+(1-f(\widetilde{E}_{\eta \alpha {\bf k}}))\ln (1-f(\widetilde{E}_{\eta
\alpha {\bf k}}))\}.  \label{entropy-h}
\end{eqnarray}
Calculating $F$ as a function of $T,H$ and $e_{B}$ enables us to find the
influence of the magnetic field on the elastic constants Eq.(\ref{el-const}).

In the heavy-fermion state the magnetization of the system per unit cell can
be given as either a sum over magnetic moments of conduction band and $f$
holes or a sum over magnetic moments of heavy quasiparticles in the hybrid
bands Eq.(\ref{h-bands}):
\begin{eqnarray}
M &=&g\mu _{B}N_{u}^{-1}\sum_{n}(<S_{cn}^{z}>+<S_{fn}^{z}>)  \nonumber \\
&=&g\mu _{B}\sum_{\alpha }\sum_{\eta =1,2}\alpha \int d\varepsilon \rho
(\varepsilon )f(\widetilde{E}_{\eta \alpha {\bf k}}).  \label{magnetization}
\end{eqnarray}
Differentiating $M$ with respect to $H$ at $H=0$ one can find the static
magnetic susceptibility per $f$ ion:\
\begin{equation}
\chi (T)=-\frac{1}{3}(g\mu _{B})^{2}j(j+1)N\sum_{\eta =1,2}\int d\varepsilon
\rho (\varepsilon )f^{\prime }(\widetilde{E}_{\eta {\bf k}}),
\label{suscept}
\end{equation}
where $f^{\prime }(E)=\partial f(E)/\partial E$. Moreover, we took into
account that at small magnetic fields the parameters $\widetilde{\varepsilon
}_{f}(T,H)$, $\widetilde{\mu }(T,H),$ $B(T,H)$ and $e_{B}(T,H)$ have field
corrections of order of O($H^{2}$). At $T=0$ we obtain\ the well known
result (see,for example, the review\cite{NewsRead}):
\begin{equation}
\chi (T=0,H=0)=\frac{1}{3}(g\mu _{B})^{2}j(j+1)N\rho _{F}^{*}.
\label{suscept0}
\end{equation}
Here the renormalized density of state $\rho _{F}^{*}$ and the effective
quasiparticle mass $m^{*}$ on the Fermi surface are given by\
\begin{equation}
\frac{\rho _{F}^{*}}{\rho _{0}}=\frac{m^{*}}{m_{0}}=1+\frac{\left| B\right|
^{2}}{T_{0}^{2}}  \label{r-DOS}
\end{equation}
In the right hand side of Eq.(\ref{r-DOS}) the first term is contributed by
conduction holes, and the second term is contributed by $f$ holes.\cite
{NewsRead,Bickers87} It is obviously that in the case $m^{*}/m_{0}>>1$ the
zero temperature susceptibility (\ref{suscept0}) is mainly determined by $f$
holes. In the case $\mu _{0}>>T_{0}$ at $T=H=0$ from Eqs.(\ref{eq1n})-(\ref
{eq2n}) we obtain a useful relation\
\begin{equation}
\left| B\right| ^{2}=\frac{N_{f}T_{0}}{N\rho _{0}}.  \label{b(T=0)}
\end{equation}
An additional assumption $N\rho _{0}T_{0}<<1$ gives the following result:%
\cite{NewsRead}\
\begin{equation}
\chi (T=0,H=0)=\frac{1}{3}(g\mu _{B})^{2}j(j+1)\frac{N_{f}}{T_{0}}.
\label{suscept00}
\end{equation}
However, the inequality $\mu _{0}>>T_{0}$ can be invalid for a semimetal
similar to YbInCu$_{4}$ having a low charge carrier concentration and large
enough $T_{0}.$ Therefore in general case it is better to use Eqs.(\ref
{suscept0}) and (\ref{r-DOS}). If $T_{0}>>T_{v}$ where $T_{v}$ is the
critical temperature of the valence phase transition, then at $T<T_{v}$ in
the HF state the susceptibility $\chi (T,H=0)$ will have a weak temperature
dependence, as a temperature correction to $\chi (0)$ is of order of O($%
T^{2}/T_{0}^{2}$).

Above $T_{v}$, that is in the normal paramagnetic state with incoherent
Kondo scattering, the total susceptibility of the electron system considered
is mainly determined by localized $f$ spins, as at temperatures $T<<\mu $
the Pauli susceptibility of conduction holes is much smaller than the
susceptibility of the localized $f$ holes weakly interacting with conduction
holes. Within the mean-field approach the interaction is neglected, and $f$
holes behave as free paramagnetic spins with the Curie-Weiss
susceptibility:\
\begin{equation}
\chi (T,H=0)=\frac{1}{3}(g\mu _{B})^{2}j(j+1)\frac{N_{0f}}{T}.
\label{suscept-p}
\end{equation}
This result is valid in the leading order in $1/N$ . In order to find
logarithmic corrections into $\chi $ due to the Kondo screening, it is
necessary to study local fluctuations of slave bosons $b_{i}$ around the
paramagnetic state.\cite{Bickers87}

One can conclude that on decreasing the temperature below $T_{v}$ the
susceptibility undergoes a jump from the value Eq.(\ref{suscept-p}) to much
the lower value Eq.(\ref{suscept0}). The result is in qualitative agreement
with experimental data (see paper\cite{FelnerNowik88} and recent data\cite
{Sarrao98,Sarrao96}).

\section{Results of numerical calculations}

In the framework of the mean-field approach to the extended lattice Anderson
model at a given temperature $T$ and pressure $P$ the free energy is
completely determined by four physical parameters: the chemical potential $%
\mu ,$ the effective energy $\varepsilon _{f}$ of the $f$ level, the order
parameter (effective hybridization) $B$ and the volume strain $e_{B}$. In
order to find the parameters we solved numerically the set of equations (\ref
{eq1n})-(\ref{eq3n}) and (\ref{pressure}) with the exchange energy Eq.(\ref
{coupling3a}). For the sake of simplicity we used a flat conduction band
with an energy independent DOS $\rho _{0}.$

We found that, depending on the $f$ level energy $E_{f}$, the hybridization
parameter $V$, the conduction band DOS $\rho _{0}$, the initial
concentration $N_{0c}$ of conduction band holes and the interaction $U_{cf}$%
, the set of the mean-field equations (\ref{eq1n})-(\ref{eq3n}) and (\ref
{pressure}) has different solutions that give different scenarios of
temperature behavior.

At first let us briefly discuss different solutions in the simplest case
when the electron-lattice interaction is neglected and $P=0$. At $U_{cf}=0$
and $T>T_{k}$ the mean-field equations has only a trivial solution with $%
B=0. $ At $T<T_{k}$ apart from the trivial solution there is one nontrivial
solution with $B\neq 0$, namely, the conventional solution describing a
continuous transition into the HF state$.$ The temperature dependence $B(T)$
of the conventional solution is represented in Fig.\ref{fig1} at parameters $%
N_{0c}=0.069,$ $\left| V/\mu _{0}\right| ^{2}=19$ and $1/N\rho _{0}J_{0}=3$
. At $T<T_{k}$ the HF state with the $B(T)$ has a lower free energy than the
trivial solution. Solving numerically Eqs.(\ref{eq1n})-(\ref{eq3n}), we have
found that at $T=0$ and $U_{cf}=0$ within the conventional solution the
valence change $\Delta N_{f}$ can achieve a maximum value of order 0.05$.$

With increasing $U_{cf}$ at fixed parameters $E_f,$ $V$, $\rho _0$ and $%
N_{0c}\sim 0.07$ the behavior of $\Delta N_f$ $(T)$ remains qualitatively
the same as for the conventional solution with $U_{cf}=0.$ There are only
quantitative differences which are nevertheless very important. At a certain
value of $U_{cf}$ $(U_{cf}/(N\rho _0\left| V\right| ^2)\sim 3)$ the valence
change $\Delta N_f(0)$ can achieve a value of order 0.1. Such a scenario we
shall use below for describing the physical properties of YbAgCu$_4.$

At large enough $U_{cf}$ $(U_{cf}/(N\rho _{0}\left| V\right| ^{2})\gtrsim 5$
and $N_{c0}\sim 0.07)$ in a certain temperature region the mean-field
equations (\ref{eq1n})-(\ref{eq3n}) have two nontrivial solutions
corresponding to minimums of the free energy $F$ as a function of the
effective hybridization $B$ (there are also solutions corresponding to
maximums of $F(T,B)$). At parameters $U_{cf}/(N\rho _{0}\left| V\right|
^{2})=5.6,$ $N_{0c}=0.069,$ $\left| V/\mu _{0}\right| ^{2}=19$ and $1/N\rho
_{0}J_{0}=3$ (below we shall use these parameters for studying properties of
YbInCu$_{4})$ these solutions are represented in Fig.\ref{fig1}. One
solution with small $B$ begins at $T=T_{k}$ , i.e. $B(T_{k})=0,$ and ends at
a point $s$ at temperature $T_{s}=0.83T_{k}$. We shall call the solution
``soft''. Near $T_{k}$ the soft solution behaves similar to the conventional
solution. Moreover, there is another solution with larger $B.$ It starts at
a point $h$ at temperature $T_{h}\approx 1.06T_{k},$ exists up to $T=0$ and
has a weak temperature dependence. We call the solution ``hard''. The free
energy $F(T,B)$ as a function of $B$ at different temperatures $T$ is
represented in Fig.\ref{fig2}. At $T>T_{h}$ the $F$ has only one minimum at $%
B=0.$ In the range $T_{k}<T<T_{h}$ the $F$ has two minimums, one at $B=0$
and the other at $B$ corresponding to the hard solution. At $T<T_{k}$ the
trivial solution gives a maximum of $F$. In the range $T_{s}<T<T_{k}$ the $F$
has two minimums related to the soft and hard solutions. Below $T_{s}$ there
is only one minimum given by the hard solution. For the given parameters a
first-order phase transition takes place at $T_{v}\approx 0.97T_{k}$ . At $%
T>T_{v}$ either soft or trivial solutions give an absolute minimum of $F$ ,
while at $T<T_{v}$ the absolute minimum is given by the hard solution. The
valence jump $\Delta N_{f}$ is of order 0.2. This scenario will be used
below for describing the first-order valence phase transition in YbInCu$_{4}$%
.

At even larger $U_{cf}$ the order parameter $B$ corresponding to the hard
solution can be much larger than $B$ of the soft solution. The hard solution
can arises well above $T_{k}$ since in this case $T_{h}>>T_{k}.$ The soft
solution can exist down to very small temperatures since in this case $%
T_{s}<<T_{k}.$ Besides, a first-order valence phase transition can occur at
a critical temperature $T_{v}>>T_{k}$ with the valence change $\Delta
N_{f}(0)\sim 1$. Such a scenario occurs, for example, for the model
parameters $U_{cf}/(N\rho _{0}\left| V\right| ^{2})=10,$ $N_{0c}=0.069,$ $%
\left| V/\mu _{0}\right| ^{2}=20$ and $1/N\rho _{0}J_{0}=4$ which give $%
T_{v}/T_{k}=130.$

Our analysis of the numerical solutions reveals that there is a correlation
between $\Delta N_f(0),$ $T_k$ and $T_v.$ If $T_k<<T_v$ , which takes place
at sufficiently large $U_{cf}$ ($U_{cf}/(N\rho _0\left| V\right| ^2)\gtrsim
9),$ then $\Delta N_f(0)$ is close to 1. If $T_v$ is only slightly larger
than $T_k,$ then $\Delta N_f(0)\sim 0.25.$ In the frame of the first-order
phase transition scenario a minimum value $\Delta N_f(0)\sim 0.2$ is
achieved when $T_v$ is slightly smaller than $T_k$.

There are also regions in the parameter space ($E_f,$ $V$, $\rho
_0,N_{0c},U_{cf}$) in which even at $U_{cf}\neq 0$ a first-order phase
transition does not occur. If the electron-lattice interaction is included
then one can expect a rich $T-P-H$ phase diagram.

\subsection{ Physical properties of YbInCu$_{4}$}

Let us apply the numerical calculations discussed above for studying the
thermodynamic properties of YbInCu$_{4}$. For the sake of simplicity we
suppose the conduction band to be flat and the degeneracy $N=8$. In
accordance with the Hall measurements \cite{Cornelius97} we use the initial
charge carrier concentration $N_{0c}$=0.069 per formula unit. The best fit
to the experimental data is found at the following parameters: $V=0.267$ eV,
$U_{cf}=0.451$ eV, $N\rho _{0}$=1.125 eV$^{-1}$, $\mu _{0}-E_{f}=0.272$ eV.
The parameter $N\rho _{0}V^{2}$ characterizing the $f$ level broadening due
to the hybridization is calculated to be 0.08 eV. One can note that for rare
earth compounds the broadening is typically of order 0.01-0.1 eV.\cite
{Lawrence} The initial chemical potential $\mu _{0}$ is calculated to be
0.061 eV. The obtained value of $V$ is in good agreement with the mixing
terms 0.27 eV between Yb 4$f$ and the Cu $p$ states found in the electronic
band structure calculations \cite{TakegaharaKasuya} of YbInCu$_{4}$.
According to the calculations \cite{TakegaharaKasuya}, the Cu $p$ states
give the main contribution into the $\Gamma _{1}$ states in the vicinity of
the Fermi surface.

The energy of the 4$f$ level, $\mu_0-E_f=0.272$ eV, is consistent with the
experimental results of PES , 0.3 eV, for the compound.\cite{Ogawa} One can
note that in other Yb based mixed valent compounds such as YbAl$_2$ and YbAl$%
_3$ the 4$f$ level energy estimated from photoemission spectra \cite
{Kaindl,Patthey} also is not centered at the Fermi energy, but 0.24 eV below
it.

For describing electron-lattice interactions it is necessary to set the
parameters $r$ and $d$. In many HF compounds the parameter $r$ is of order $%
-8<r<-2.$ We take $r=-2$ and $d=2$ eV that give $\Omega _{1}\approx 12$ and $%
\Omega _{2}=-25$ in Eq.(\ref{Grueneisen}). The unit cell of YbInCu$_{4}$
contains four Yb ions and has volume 372$\times 10^{-24}$cm$^{3}.$ Therefore
each Yb ion occupies volume $v_{0}=$93 $\times 10^{-24}$cm$^{3}.$ Since
according to experimental data\cite{Kindler,Teresa} in the normal state the
bulk modulus of YbInCu$_{4}$ is equal to $11,1\times 10^{11}$ erg/cm$^{3},$
the parameter $C_{B}$ in Eq.(\ref{lat-energy}) is taken to be equal to $%
c_{B}\nu _{0}=1,03\times 10^{-10}$ erg.

Our analysis of the numerical solutions of the mean-field equations (\ref
{eq1n})-(\ref{eq3n}) and (\ref{pressure}) shows that for the parameters
chosen the system under consideration undergoes a first-order isostructural
valence phase transition at the critical temperature $T_{v}=$42 K. Above $%
T_{v}$ the system is in the normal state. Below $T_{v}$ the HF state
described by the hard solution is formed, as the free energy of the HF state
becomes smaller than the free energy of the normal state with the incoherent
Kondo scattering. It is demonstrated by Fig.\ref{fig3}. Within the
mean-field solution at $T>T_{v}$ the Yb ions have the integral valency +3
corresponding to $B=0.$ At $T<T_{v}$ the Yb ions have a non-integral valence
$3-\Delta N_{f}$ . A temperature dependence of the $\Delta N_{f}$ is shown
in Fig.\ref{fig4}. At $T=T_{v}$ there is a jump of the valence, $\Delta
N_{f}=0.24.$ With decreasing temperature the valence change $\Delta N_{f}$
achieves the value 0.27. The value is larger than 0.17 deduced from $L_{3}$
measurements. \cite{Cornelius97} Reasons of the discrepancy will be
discussed below.

\subsubsection{Thermal properties of YbInCu$_4$}

In Fig.\ref{fig5} we represent the temperature behavior of the entropy ($S$)
of the system considered. In the framework of our approach above $T_{v}$ the
entropy is large enough due to a large contribution given by localized $f$
holes. At $T_{v}$ the entropy drops from $S(T_{v}+0)=25$ J/mol K to $%
S(T_{v}-0)=4.9$ J/mol K$,$ i.e. $\Delta S=20$ J/mol K$.$ These theoretical
estimates are larger than experimental data: $\Delta S=10$ J/mol K, $%
S(T_{v}+0)=13$ J/mol K$,$ $S(T_{v}-0)=3$ J/mol K$.$ \cite{Sarrao98} However,
the theoretical estimate of the ratio $S(T_{v}+0)$/$S(T_{v}-0)=5.1$ is in
good agreement with the experimental value $S(T_{v}+0)$/$S(T_{v}-0)=4.3$.
Below $T_{v}$ $f$ holes are strongly hybridized with conduction band holes
and form the nonmagnetic heavy fermion Fermi-liquid ground state. That is
why the entropy tends to zero in the limit $T\rightarrow 0.$

There are a few reasons of the divergences between our estimates and
experimental data.. At first, we neglected a crystal field splitting of the
8-fold degenerate $f$ level. Inelastic neutron measurements \cite{Severing}
revealed that the quartet $\Gamma _8$ is the ground state, but the splitting
$\Delta _c$ is quite small, $\Delta _c=32$ K. The experimental value $%
S(T_v+0)=13$ J/mol K$\approx R\ln 5$ is close to the value $R\ln 4$ expected
for four-fold degenerate $\Gamma _8$ state. An account of the crystal-field
splitting within our approach will result in decreasing entropy and a more
remarkable temperature dependence of the entropy above $T_v$. At second, the
mean field approach does not give a correct result of the entropy of the
localized $f$ electrons in the high temperature phase. Namely, the approach
results in the value $S=R(\ln N-(N-1)\ln (1-N^{-1}))$ for the entropy of
noninteracting $f$ holes instead of the value $R\ln N$. To obtain a correct
result it is necessary to go beyond the mean-field solution and take into
account fluctuation corrections in next orders in $1/N$.

In the HF phase at $T<<T_{v}$ the linear coefficient of the specific heat $%
\gamma $ is given by the relation $\gamma =\frac{1}{3}\pi ^{2}k_{B}^{2}N\rho
_{F}^{*}$ where $\rho _{F}^{*}$ is determined by Eq.(\ref{r-DOS}). For the
parameters chosen we find $\gamma =71$ mJ/mol K$^{2}.$ The experiment \cite
{Sarrao98} gives slightly smaller value $\gamma \approx 50$ mJ/mol K$^{2}.$

\subsubsection{Magnetic susceptibility of YbInCu$_4$}

Using Eq.(\ref{suscept}) we calculated the temperature dependence of the
spin susceptibility $\chi (T)$. Corresponding results are represented in Fig.%
\ref{fig6}. Above $T_{v}$ localized $f$ electrons give the main contribution
into the susceptibility. As a result, the $\chi $ follows the Curie-Weiss
law: $\chi =C/T$. Below $T_{v}$ hybrid quasiparticles are formed, and the
electron system under consideration behaves as a Fermi-liquid system with an
enhanced Pauli susceptibility. At zero temperature the $\chi (0)$ is given
by Eq.(\ref{suscept0}). At ambient pressure we found $T_{L}\equiv C/\chi
(0)=384$ K. The experimental value is 470 K. \cite{Cornelius97} We
calculated also the low temperature Kondo scale $T_{0}$ defined by Eq.(\ref
{T0}) and found $T_{0}=315$ K. In the conventional HF systems it is the $%
T_{0}$ that plays the role of a universal energy scale for low temperature
thermodynamic and transport phenomena in the HF state (see Refs.\onlinecite
{NewsRead} and \onlinecite{MillisLee}, for example). According to Eq.(\ref
{suscept00}), in the limit $\mu _{0}>>T_{0}$ there is a simple relation
between $T_{0}$ and $T_{L}:$ $T_{0}=N_{f}$ $T_{L}=(1-\Delta N_{f})$ $T_{L}.$
The approximate relation at $\Delta N_{f}=0.27$ gives $T_{L}$= 430 K that is
slightly larger than our calculated value ($T_{L}=384$ K). The difference is
a consequence of the approximate character of the relation$,$ because in our
analysis we have $\mu _{0}/T_{0}\sim 2.2$ due to the small hole
concentration $N_{0c}=0.069$ per $f$ ion. Our calculations revealed that
both the $T_{L}=384$ K and $T_{0}=315$ K are much larger than the Kondo
temperature $T_{k}=40.4$ K characterizing Kondo effect at $T>T_{v}$. The
result is in agreement to inelastic neutron scattering measurements \cite
{Severing,Lawrence97,Lawrence99} that also revealed an enhancement of the
energy scale from $T_{k}=25$ K to $T_{0}=405$ K. The very large enhancement
of $T_{0}$ in comparison with $T_{k}$ is brought about by the enhancement of
the exchange energy $J$. In papers\cite{Cornelius97,Figuer} in order to
explain the enhancement of $T_{L}$ it was suggested that the effect is
caused by a strong energy dependence of the conduction band DOS $\rho
(\varepsilon ).$ Namely, $\rho (\varepsilon _{F})$ must be sharply
increasing function of the Fermi energy $\varepsilon _{F}$. The results of
our calculations show that in terms of the extended lattice Anderson model
the enhancement can be explained even in the case of a flat conduction band.

The enhancement of $\gamma $ and $\chi (0)$ with respect to the values
typical for normal metals is explained by the increase of the DOS on the
Fermi surface. According to Eq.(\ref{r-DOS}), there is also a quasiparticle
mass enhancement. Our calculations give $m^{*}/m_{0}=27.$ Therefore, one can
classify YbInCu$_{4}$ as a ``light'' heavy-fermion compound as it was
suggested by Felner et al.\cite{FelnerNowik87}

\subsubsection{Volume expansion and bulk modulus in YbInCu$_4$}

In Fig.\ref{fig7} we represent results of the numerical calculations of the
volume strain $e_{B}=\Delta v/v.$ In good agreement with experimental data
\cite{FelnerNowik87,Cornelius97,Teresa} we find a sharp volume expansion
about 0.5\% when decreasing $T$ below $T_{v}$. Calculating the volume strain
at different pressures, we estimated the isothermal compressibility $%
k_{T}=-d\ln v/dP$ and found $k_{T}(T=20$ K$)=1.05$ Mbar$^{-1}.$ Experimental
measurements \cite{Teresa} gave $k_{T}(T=20$ K$)=1.2$ Mbar$^{-1}.$
Ultrasonic studies \cite{Kindler,Zherlitsyn} resulted in $k_{T}(T=20$ K$%
)=0.95$ Mbar$^{-1}$.

Using Eq.(\ref{el-const}) we calculated the temperature dependence of the
bulk modulus $c_{B}=1/$ $k_{T}.$ These results are presented in Fig.\ref
{fig8}. Such a temperature behavior is in complete agreement with the
experimental data. \cite{Kindler,Zherlitsyn}

According to Eq.(\ref{strain-g}), at a sufficiently small electron-lattice
coupling the volume strain must be proportional to $\Delta N_{f}$: $%
e_{B}=a\Delta N_{f}.$ Our calculations conform the relation and give $a$%
=0.019. The experimental estimation of the linear coefficient $a$ is 0.046.
\cite{Cornelius97}

\subsubsection{Pressure effect in YbInCu$_4$}

The effect of pressure on the physical properties of the system under
consideration is represented in Fig.\ref{fig6}. According to our
calculations, applying pressure shifts the critical temperature $T_{v}$ of
the first- order phase transition to lower temperatures at the rate $%
dT_{v}/dP=-0.13$ K kbar$^{-1}$. The theoretical estimate is one order of
magnitude smaller than the experimental result\cite
{FelnerNowik88,Sarrao98,Teresa} $,-2.2$ K kbar$^{-1}$. In order to improve
the result, it is necessary probably to go beyond the mean-field theory.
Pressure results in decreasing the valence change at the rate $d\Delta
N_{f}(3.5$ K)$/dP=-6.3$ Mbar$^{-1\text{ }}$in a satisfactory agreement with
the experimental value $-4.5$ Mbar$^{-1\text{ }}$ found from volume
expansion measurements.\cite{Teresa} In Fig.\ref{fig6} we plot magnetic
susceptibility at ambient pressure and $P=3$ kbar. From the results we find $%
dT_{L}/dP=-12.7$ K/kbar in good agreement with the experimental value $-12.4$
K/kbar obtained by Sarrao et al.\cite{Sarrao98} The theoretical estimate of
the Gr\"{u}neisen parameter $\Gamma =-d\ln T_{L}/d\ln v$ $=-31$ is in
excellent agreement with the experimental value\cite{Sarrao98} $\Gamma
=-30.6 $. The physical origin of the pressure effect is related mainly to
the influence of pressure on fluctuations of the $f$ shell size. Indeed, an
applied pressure decreases volume, $e_{B}<0$. In turn, this results in
decreasing the exchange interaction $J$ (see Eq.(\ref{coupling3a})) and,
consequently, decreasing the Kondo scales $T_{k}$ and $T_{0}$. We believe
that the $f$-shell-size-fluctuation mechanism explains also the negative $%
\Gamma $ observed in other Yb based HF compounds and their pressure
dependent properties (see, for example, Ref.\cite{Abd} and reference
therein).

\subsubsection{$H-T$ phase diagram.}

Basing on the results of Sec. V we investigated the effect of a magnetic
field on the first-order valence phase transition. Only the Zeeman energy
was taken into account in the numerical calculations. In our calculations we
took the gyromagnetic factor $g=8/7$ expected for $j=7/2.$ For the sake of
simplicity we neglected the electron-lattice interaction, as it gives only
about 10\% correction in accordance to our estimates. Solving
selfconsistently Eqs.(\ref{eq1h})-(\ref{eq3h}) we found that magnetic field
pushes the first-order valence phase transition to lower temperatures. The
obtained $T-H$ phase diagram is represented in Fig.\ref{fig9}. It is
interesting to note that the elliptic equation\
\begin{equation}
\left( \frac{H_{v}(T)}{H_{v}(T=0)}\right) ^{2}+\left( \frac{T_{v}(H)}{%
T_{v}(H=0)}\right) ^{2}=1  \label{h-t-line}
\end{equation}
is a good fit to the critical line of the first-order phase transition $%
H_{v}(T)$ given by our numerical calculations. Recently the relationship Eq.(%
\ref{h-t-line}) between $H_{v}(T)$ and $T_{v}(H)$ was inferred from
magnetoresistance measurements.\cite{Immer} Our estimate of the critical
field $H_{v}(T=0)=47$ T is also in a satisfactory agreement with
experimental value $H_{v}(T=0)=34$ T. The critical line $H_{v}(T)$ divides $%
H-T$ plane into the low temperature-low magnetic field region in which the
system is in the mixed-valence heavy fermion ground state, and the high
temperature-high magnetic field region in which the system is in a normal
state with stable magnetic moments of Yb ions.

\subsubsection{Electrical resistivity and the Hall effect in YbInCu$_{4}$}

In accordance to the measurements \cite{Nakamura94,Sarrao96} the electrical
resistivity $R(T)$ in YbInCu$_{4}$ demonstrate two peculiarities: (i) a
large drop at $T=T_{v}$ and (ii) a weak temperature dependence at $T<T_{v}$.
Analyzing the data at $T>T_{v}$, one can distinguish two main contributions
into the resistivity: a large residual resistivity $R_{0}$ and a linear
temperature dependent contribution given by electron scattering off phonons.
At $T=T_{v}$ the phonon contribution is much smaller than $R_{0}$ and as a
result $R(T_{v}+0)\approx R_{0}$. In the low temperature phase the residual
resistivity $R_{0}^{*}$ is approximately 12 times smaller than $R_{0}$.

Basing on our approach let us analyze the temperature behavior of the
resistivity at $T<T_v$. At first, one can note that in the HF state the
renormalized electron-phonon interaction is very small and its contribution
into resistivity can be neglected (see, for example, Ref.\onlinecite{grewe}%
). Only impurity scattering and collisions between heavy fermions determine
the resistivity. It leads to $R^{*}(T)=R_0^{*}+A^{*}T^2$, where the
coefficient $A^{*}\sim T_0^{-2}$. \cite{Coleman87b,Auerbach,MillisLee}.
Since $T{_0}>>T_v$, the temperature dependence of $R^{*}(T)$ must be weak
enough, which is in agreement with the experimental data.\cite
{Nakamura94,Sarrao96}

Let us estimate and compare the residual resistivity above and below $T_{v}$%
. At $T>T_{v}$ the residual conductivity can be estimated using the well
known relation
\begin{equation}
\sigma _{0}\sim e^{2}\rho _{0F}v_{0F}^{2}\tau _{0}\sim e^{2}N_{0c}\tau _{0}
\label{conduct}
\end{equation}
where $v_{0F}=p_{0F}/m_{0}$ is the Fermi velocity, $p_{0F}$ is the Fermi
momentum, $\rho _{0F}\sim p_{0F}m_{0}$ is the conduction band DOS, $\tau
_{0}^{-1}$ is a rate of a potential scattering of charge carriers off
impurities and lattice imperfections. Here we also used that $N_{0c}\sim
p_{0F}^{3}$. In the HF state the residual conductivity is determined by a
potential scattering of heavy fermions and is given by a similar equation
with the replacement of $\rho _{0F}$, $v_{0F}$, $p_{0F}$ and $\tau _{0}$ by
the renormalized parameters $\rho _{F}^{*}\sim p_{F}m^{*}$, $%
v_{F}=p_{F}/m^{*}$ and $\tau ^{*}=\tau _{0}m^{*}/m_{0}.$\cite
{MillisLee,Fukuyama} Moreover, it is necessary to take into account that the
renormalized Fermi surface determined by the total number of charge carriers
$N_{t}=N_{0c}+N_{0f}$, that is (($p_{F}/p_{0F})^{3}=N_{t}/N_{0c}$. Then, we
find the residual conductivity at $T<T_{v}:$%
\begin{equation}
\sigma ^{*}\sim e^{2}\rho _{F}^{*}(v_{F})^{2}\tau _{0}^{*}\sim
e^{2}N_{t}\tau _{0}.  \label{hfcond}
\end{equation}
Therefore, we have $R_{0}/R_{0}^{*}\simeq N_{t}/N_{0c}$. Above we have found
$N_{t}=1.07$ while $N_{0c}$=0.07. It gives the crude theoretical estimation $%
R_{0}/R_{0}^{*}\simeq 15$ in agreement with the experimental value \cite
{Sarrao96}, $R_{0}/R_{0}^{*}\simeq 12$. Thus, we can conclude that the drop
of the resistivity and change of its temperature behavior when transiting
into the low temperature mixed-valent state in YbInCu$_{4}$ are mainly
related to (i) the decrease of the residual resistivity due to increase of
the charge carrier concentration and (ii) the suppression of the
electron-phonon scattering.

Basing on our approach one can explain also the jump of the charge carrier
concentration at $T=T_{v}$ observed in the Hall measurements$.$\cite
{Cornelius97,Figuer} Indeed, at $T>T_{v}$ $f$ electrons are localized and
only conduction band holes with the concentration $N_{0c}$ participate in
transport phenomena and the Hall effect. At $T<T_{v}$ the $f$ electron
states are hybridized with conduction band states, which results in the
formation of hybrid quasiparticles (heavy fermions). The total number of the
quasiparticles under the renormalized Fermi surface is equal to $%
N_{t}>N_{0c} $. All these quasiparticles give a contribution into the Hall
constant.

\subsection{Physical properties of YbAgCu$_4$}

As we have discussed above, with decreasing the interaction $U_{cf}$ the
system under consideration reveals a crossover from the first-order valence
phase transition to a continuous formation of the HF state. We use such a
scenario for describing thermodynamic properties of YbAgCu$_{4}$. In many
respects YbAgCu$_{4}$ is close to YbInCu$_{4}$, but YbAgCu$_{4}$ has a
normal metallic charge carrier concentration. For our calculations we take $%
N_{0c}$=0.21 per Yb, which corresponds to 0.84 charge carrier per unit cell.
The best fitting to the experimental data is found at the following
parameters: $V=0.24$ eV, $1/N\rho _{0}$=0.9 eV, $\mu _{0}-E_{f}$=0.25 eV
which are close to ones for YbInCu$_{4}$. Only the interaction $U_{cf}=0.276$
eV is taken almost twice smaller than $U_{cf}$ in YbInCu$_{4}$. The small
value of $U_{cf}$ can be explained by a stronger charge screening due to a
larger charge carrier concentration in comparison to YbInCu$_{4}$. For these
parameters we have $N\rho _{0}V^{2}=$0.064 eV, $\mu _{0}=0.19$ eV. The value
of 4$f$ level energy is consistent with the result of PES \cite
{Weibel,Malterre}, 0.3 eV.

For describing the electron-lattice interactions in YbAgCu$_{4}$ we use the
same $r$ and $d$ as for YbInCu$_{4}$, i.e. $r=-2,$ $d=2$ eV, and $%
C_{B}=1,03\times 10^{-10}$ erg. For the parameters chosen the equations (\ref
{eq1n})-(\ref{eq3n}) and (\ref{pressure}) have only one nontrivial
(conventional-like) solution below $T_{k}=86$ K. Results of numerical
calculations based on the solution are represented in Figs.\ref{fig3}-\ref
{fig8}.

Let us compare temperature behavior of YbAgCu$_{4}$ and YbInCu$_{4}$. In
accordance to our numerical calculations the HF state in YbAgCu$_{4}$ is
continuously formed below $T_{k}=86$ K. This temperature correlates with the
temperature of the resistance maximum, $T_{max}\approx 75-90$ K. \cite
{Sarrao96} According to our calculations represented in Fig.\ref{fig4}, at $%
T\rightarrow 0$ the valence change $\Delta N_{f}$ tends to 0.13 in very good
agreement with the $L_{3}$ data\cite{Cornelius97}$.$

Temperature behavior of the entropy is represented in Fig.\ref{fig5}.
Unfortunately, entropy measurements in YbAgCu$_{4}$ are unknown for us to be
compared with our calculations. The calculated value of the linear
coefficient of the specific heat $\gamma =230$ mJ/mol K$^{2}$ is between the
data\cite{Cornelius97} $\gamma =220$ mJ/mol K$^{2},$ and $\gamma =250$
mJ/mol K$^{2}$ found from other measurements.\cite{Sarrao96}

The calculated magnetic susceptibility $\chi (T)$ represented in Fig.\ref
{fig6} has a broad maximum at $T\approx 40$ K and its maximum value is
approximately twice smaller than $\chi (T=T_{v})$ in YbInCu$_{4}$, which is
in good agreement with the magnetic measurements.\cite{Sarrao96} The
calculated value $T_{L}=120$ K is slightly smaller than the experimental data%
\cite{Cornelius97,Sarrao96} $T_{L}=150$ K.

The calculated mass enhancement in YbAgCu$_{4}$ is equal to $m^{*}/m_{0}=87$
. This value is in three times larger then the mass enhancement in YbInCu$%
_{4}.$ Therefore, one can classify YbAgCu$_{4}$ as a ``moderate''
heavy-fermion compound.

The temperature dependence of the volume strain $e_{B}$ is shown in Fig.\ref
{fig7}. In YbAgCu$_{4}$ the zero temperature value $e_{B}(T=0)=0.24$ \% is
approximately twice smaller than the $e_{B}(0)$ in YbInCu$_{4}.$ We find
that the relation $e_{B}=a\Delta N_{f}$ holds also in YbAgCu$_{4}$ with $%
a=0.018$.

Using the parameters given above we calculated the temperature behavior of
the bulk modulus for two concentrations of charge carriers, $N_{0c}$=0.21
and $N_{0c}$=0.07. In the both cases the calculations revealed a minimum of
the bulk modulus. These results are represented in Fig.8. One can see that
with increasing the charge carrier concentration the minimum becomes smaller
and less expressive. The results of our calculation for $N_{0c}$=0.07 can be
related to the behavior of the bulk modulus of YbIn$_{1-x}$Ag$_{x}$Cu$_{4}$
at $x\approx 0.3$, \cite{Zherlitsyn} because in the case the charge carrier
concentration is still small but the system already shows a continuous
transition into the HF state. \cite{Sarrao96} The calculated behavior of the
bulk modulus represented in Fig.\ref{fig8} is in good agreement, both
qualitative and quantitative, with the acoustic measurements. \cite
{Zherlitsyn}

\section{Discussion and conclusions}

In the present section we shall discuss the mean-field approximation used in
our paper and other problems related to physical phenomena in YbIn$_{1-x}$Ag$%
_{x}$Cu$_{4}$, and summarize results of our numerical calculations performed
in the framework of the extended lattice Anderson model proposed in the
present paper.

Our consideration of the model was based on the mean-field approximation
that corresponds to taking into account the leading order of the $1/N$
expansion. The mean-field solution is the exact solution of the extended
lattice Anderson model in the limit $N\rightarrow \infty .$ In the case
under consideration the degeneracy $N$ of the electron state 4$f^{13}$ is
finite and equal to 8. Therefore, we face the problem of fluctuations around
the mean-field solution. The fluctuations give $1/N$ corrections to the
solution. Unfortunately, so far there is no a detail analysis of the
corrections into thermodynamic properties of the lattice Anderson model.
There is only an exact solution of the single-impurity Anderson model with
Hamiltonian Eq.(\ref{Anderson}) for arbitrary degeneracy $N$.\cite
{ColemanAndrei} In the paper\cite{ColemanAndrei} it has been shown that in
the Fermi-liquid regime at $T<<T_{0}$ the magnetic susceptibility and heat
capacity given by the mean-field solution are in a remarkably good
quantitative agreement with the exact solution even for $N=2$. For $N=8$ the
region of a quantitative agreement becomes broader. A noticeable error
occurs only around the crossover from weak to strong coupling ($T\sim T_{k}$%
) where the critical fluctuations of the slave-boson field about the
mean-field solution are greatest. However, a qualitative agreement takes
place even in the region. With increasing temperature above $T_{k}$ the
mean-field solution converges quickly to the exact solution, as both
solutions describe a paramagnetic state with localized 4$f$ electrons that
weakly interact with conduction band electrons. As shown in \cite
{RasulHewson} at high temperatures, the impurity susceptibility approaches
Curie behavior with logarithmic corrections of order O($1/N$) that are
produced by slave-boson fluctuations around the normal state.

There is the only significant defect of the mean-field solution related to
the fact that in the limit $N\rightarrow \infty $ the crossover from weak to
strong coupling sharpens into a second order phase transition at $T_{k}$.
Besides, at $T=T_{k}$ the susceptibility has a break in the temperature
dependence.

Basing on the results let us consider the extended lattice Anderson model.
In the framework of the mean-field approach both the single-impurity and
lattice Anderson models are described by the same order parameter (see Eq.(%
\ref{B-definition})). However, unlike the single impurity model, the lattice
model contains spatial correlations of the order parameter. It is well known
from the theory of critical phenomena that with increasing the
dimensionality, order parameter fluctuations become weaker. Therefore, one
can expect that critical fluctuations in the 3D lattice model around the
mean-field solution are not stronger than the fluctuations in the
single-impurity model studied in the paper\cite{ColemanAndrei}.

From this point of view let us discuss our mean-field solutions for
compounds YbInCu$_{4}$ and YbAgCu$_{4}$. According to the experimental data%
\cite{Cornelius97,Severing,Lawrence97,Lawrence99} and our calculations, the
low temperature Kondo scale $T_{0}$ is about 400K in the low temperature
phase of YbInCu$_{4}$, that is below $T_{v}=42$ K. As $T_{v}<<T_{0},$ one
can expect that at $T<T_{v}<<T_{0}$ fluctuations around the mean-field
solution are weak and give small corrections with respect to small
parameters $1/N$ and $T/T_{0}$. We think that it explains a good agreement
between experimental data and our numerical calculations of susceptibility,
entropy, valence change of Yb ions, volume change, bulk modulus and other
physical parameters presented in Sec. VIA. At temperatures $T>T_{v}$ we are
in the regime $T>T_{v}>T_{k}=25$ K, therefore, we again expect that
corrections of order O($1/N$) due to slave-boson fluctuations around the
normal state are small and become noticeable only at $T$ close to $T_{v}$.
As at these temperatures there are no correlations between Kondo scattering
of electrons off different localized $f$ electrons, $1/N$ corrections to
thermodynamic properties are the same as in the single-impurity Anderson
model.

In YbAgCu$_{4}$ we have a similar situation. We expect that at low
temperatures $T<<T_{0}=150$ K the mean-field solution gives a good
description of thermodynamic properties. The maximum error occurs in the
crossover region with $T_{k}=86$ K. It is interesting to note that according
to our numerical calculations, in the lattice case the break of temperature
behavior of $\chi (T)$ at $T_{k}$, which is an artifact of the mean-field
approximation, is very small in comparison to the mean-field solution of the
single-impurity Anderson model\cite{ColemanAndrei}.

In order to check the mean-field solution of the lattice Anderson model Eq.(%
\ref{Anderson}) one could use the dynamical mean-field theory that at the
present time attracts much attention. \cite{Georges96} For the purpose in
terms of the theory it is necessary to solve the model Eq.(\ref{Anderson})
at different degeneracy $N=2,4,6,8...$ , and then to compare the solutions
with the mean-field solution based on the $1/N$ expansion as it has been
done for the single-impurity Anderson model\cite{ColemanAndrei}.
Unfortunately this problem is still open and is out of the scope of the
present paper.

Above we have mentioned that in accordance to the Hall measurements\cite
{Cornelius97,Figuer} in compounds YbIn$_{1-x}$Ag$_{x}$Cu$_{4}$ at $x<0.2$
the charge carrier concentration is low, 0.07 per formula unit. However,
neutron scattering\cite{Lawrence97} and susceptibility measurements\cite
{Sarrao96} can be well interpreted in terms of the single-impurity Anderson
model with a metallic electron concentration. It enables us to conclude that
in the compounds the charge carrier concentration is larger than that
concentration at which it would be necessary to take into account
Nozi\`{e}res exhaustion principle. In the case $x>0.3$ the charge carrier
concentration quickly achieves a normal metallic concentration, and the
problem of a low charge carrier concentration already does not confront
before us.

The $L_{3}$ measurements\cite{Cornelius97} in YbInCu$_{4}$ give a clear
evidence for a non-integral valence of Yb ions in the normal state. However
there is a profound difference between the origin of the non-integral
valence of Yb ions above and below $T_{v}$. In the high temperature region ($%
T>T_{k}$) the non-integral valence of Yb ions is produced by uncorrelated
electron transitions between the $f$ shell of a Yb ion and conduction band.
The transitions do not breakdown the localized character of $f$ electron
states. Certain correlations due to the Kondo effect arise only when the
temperature lowers to temperatures about $T_{k}.$ As we have shown above
within our approach, at $T<T_{v}$ the mixed valence state of Yb ions is
caused by the effective hybridization between 4$f$ and conduction band
states, which results in a renormalization of the conduction band and the
formation of hybrid quasiparticle states due to the coherent Kondo effect.
In other words, one can say that at $T<T_{v}$ the electron transitions
become strongly correlated. From this point of view the change of the Yb ion
valence is related to the change of the spatial electron distribution over
electron states of all ions, including Cu and In ions, participating in the
formation of the renormalized conduction band, which is consistent with the
NQR and Knight shift measurements \cite
{Nakamura90,Graham,Nakamura96,Nakamura98}. This physical picture is also
consistent with the theory of the mixed-valent state formation for a single
Kondo impurity in terms of the single impurity Anderson model (see, for
example,\cite{NewsRead,Schlottmann}).

There is another interesting problem related to the crystal field splitting
of the degenerate 4$f$ level in YbInCu$_{4}$. The inelastic neutron
scattering measurements\cite{Severing} have revealed a crystal-field (CF)
splitting of order 32 K in high temperature phase. Below $T_{v}$ the CF
excitations disappear. Recently the effect of the hybridization on the CF
splitting was also observed in certain Ce based compounds of type ReNi by
use of neutron spectroscopy\cite{Lazukov} at temperatures below $T_{k}$ in
the Fermi-liquid regime. Therefore, one can suppose that the hybridization
and formation of the heavy fermion ground state at $T<T_{v}$ also influence
the temperature behavior of the CF excitations in YbInCu$_{4}$.

The theory of the first order valence phase transition developed in the
present paper can be applied also for studying valence transitions in Ce
based and other rare earth compounds. Specifically, recent measurements\cite
{Murani} on Ce alloyed with 7 at.\% Sc to stabilize the $\gamma -$Ce phase
against $\beta -$Ce formation demonstrated a similar physical behavior at
the $\gamma -\alpha $ isostructural first-order phase transition as the
behavior observed in YbInCu$_{4}$, namely, localized $f$ electrons in $%
\gamma -$Ce and delocalized ones in $\alpha -$Ce, a valence change $\Delta
N_{f}\approx 0.2$, a large enhancement of the Kondo scale from $T_{k}\sim 60$
K in $\gamma -$Ce to $T_{0}\sim 1800$ K in $\alpha -$Ce and so on. However,
according to experimental data, the electron-lattice coupling in Ce based
compounds is stronger than this coupling in Yb compounds and can be a
driving force of the transition.\cite{Allen82,Allen92}

In summary, we have used the extended lattice Anderson model for
investigating the first-order valence phase transition in YbInCu$_{4}.$ The
model takes into account Coulomb repulsion $U_{cf}$ between $f$ and
conduction band holes, and two mechanisms of the electron-lattice
interaction. For the model we have developed a mean-field approach based on
the $1/N$ expansion method. Within the mean-field approach we have found
that the system under consideration undergoes a first-order phase transition
from the normal state into the heavy-fermion state with mixed valent $f$
ions. We have found that the Coulomb interaction $U_{cf}$ strongly
influences on the exchange interaction $J$ between spins of $f$ and
conduction band electrons. The $U_{cf}$ enhances $J$ and, as a result, the
coherent Kondo effect is also enhanced. It is the driving force of the
first-order valence phase transition. Basing on the model, we have carried
out numerical calculations of temperature and pressure dependences of
thermal, magnetic and elastic properties of YbInCu$_{4}$ and obtained a good
agreement with experimental results. We have studied the role of the
electron-lattice coupling in the compound and found that the $f$
shell-size-fluctuation mechanism of the electron-lattice interaction gives
the main contribution into the volume change of the lattice and determines
the negative sign of the Gr\"{u}neisen parameter. Our analysis have shown
that the evolution of the first-order phase transition into a continuous
transition in the series of compounds YbIn$_{1-x}$Ag$_{x}$Cu$_{4}$ can be
explained by decreasing the interaction $U_{cf}$ with increasing Ag
concentration $x$ due to increase of the charge carrier concentration and an
enhancement of charge screening. In the framework of the extended Anderson
model we have found that a magnetic field pushes the valence transition to
lower energies, and in the $H-T$ plane the critical line of the first-order
valence phase transition is described by an elliptic equation in complete
agreement with experimental data

\acknowledgments

One of the authors (A.G.) gratefully acknowledges the Physics Institute of
the Frankfurt University for hospitality as well as the Russian Fund of
Fundamental Investigations for the financial support in part under Grant No.
98-02-18299.

\begin{figure}[tbp]
\caption{Order parameter (the effective hybridization) $B$ versus $T.$ The
conventional solution of the mean-field equations Eqs.(\ref{eq1n})-(\ref
{eq3n}) for the interaction $U_{cf}=0$ is given by the dotted curve. Curves $%
s$ and $h$ represent ``soft''and ``hard'' solutions of the equations,
respectively, at a moderately large $U_{cf}$ ($U_{cf}=0.451$ eV, $N=8,$ $%
V=0.267$ eV, $N\rho _{0}$=1.125 eV$^{-1},$ $N_{0c}$=0.069 per formula unit, $%
\mu _{0}-E_{f}=0.272$ eV) when $T_{k}/\mu _{0}=0.057$ and $T_{v}/\mu
_{0}=0.055$. Here the electron-lattice interaction is neglected. }
\label{fig1}
\end{figure}

\begin{figure}[tbp]
\caption{Free energy $F$ of the extended lattice Anderson model as a
function of the order parameter (the effective hybridization) $B$ at
different temperatures $T/\mu _{0}=$0.04, 0.05, 0.06, 0.07 for the same
parameters $V,N,N\rho _{0}$ and $N_{0c}$ as those in Fig.1 ($T_{k}/\mu
_{0}=0.057$, $T_{v}/\mu _{0}=0.055)$. A deep minimum arising at temperature $%
T/\mu _{0}\lesssim $0.06 and $B$ near $2$ corresponds to the ``hard''
solution. The arrow shows the position of a minimum of the $F$ corresponding
to the ``soft'' solution.}
\label{fig2}
\end{figure}

\begin{figure}[tbp]
\caption{Temperature dependence of the free energy $F$ in the normal state
(curve 1) and the heavy-fermion (HF) state (curve 2) in YbInCu$_{4}$ and
YbAgCu$_{4}.$ In YbInCu$_{4}$ the HF state is described by the hard solution
of the mean-field equations while in YbAgCu$_{4}$ it is described by the
coventional solution. The free energies are given with the accuracy up to a
certain energy shift.}
\label{fig3}
\end{figure}

\begin{figure}[tbp]
\caption{Valence change $(\Delta N_f)$ versus temperature $T$ in YbInCu$_4$
(curve 1). Curve 2 shows the $\Delta N_f$ in the case when the electron
lattice interaction is neglected. Curve 3 is the $\Delta N_f$ in YbAgCu$_4$.
Curve 4 is the $\Delta N_f$ in the case when both the electron-lattice
coupling and the on-site Coulomb repulsion $U_{cf}$ are neglected.}
\label{fig4}
\end{figure}

\begin{figure}[tbp]
\caption{Temperature dependence of the entropy $S(T)$ in YbInCu$_4$ (solid
curve) and in YbAgCu$_4$ (dotted curve).}
\label{fig5}
\end{figure}

\begin{figure}[tbp]
\caption{Magnetic susceptibility $\chi $ versus temperature $T$ in YbInCu$_4
$ at pressure $P=0$ (curve 1) and $P=3$ kbar (curve 2). Curve 3 is the $\chi
$ in YbAgCu$_4$ at $P=0$ and $T_k=86$K. Curve 4 is the $\chi $ for the same
parameters as for YbInCu$_4$ but when both the electron-lattice coupling and
the Coulomb interactions $U_{cf}$ are neglected.}
\label{fig6}
\end{figure}

\begin{figure}[tbp]
\caption{Volume strain $e_B$ against $T$ in YbInCu$_4$ (solid curve) and in
YbAgCu$_4$ (dotted curve).}
\label{fig7}
\end{figure}

\begin{figure}[tbp]
\caption{Bulk modulus $c_B$ versus $T$ in YbInCu$_4$ (solid curve), YbAgCu$%
_4 $ (dotted curve) and YbIn$_{0.7}$Ag$_{0.3}$Cu$_4$ (dash-dotted curve).}
\label{fig8}
\end{figure}

\begin{figure}[tbp]
\caption{Normalized critical magnetic field $H_{v}(T)/H_{v}(T=0)$ versus the
normalized temperature $T/T_{v}(H=0)$ in YbInCu$_{4}$. Results of the
numerical calculations are presented by solid squares. The critical line of
the first-order valence phase transitions divides the $H-T$ phase diagram of
into two regions corresponding to the mixed-valence heavy fermion (HF) state
and a normal paramagnetic (P) state with stable magnetic moments. The
elliptic equation (\ref{h-t-line}) (dotted curve) is a good fit to the
critical line.}
\label{fig9}
\end{figure}

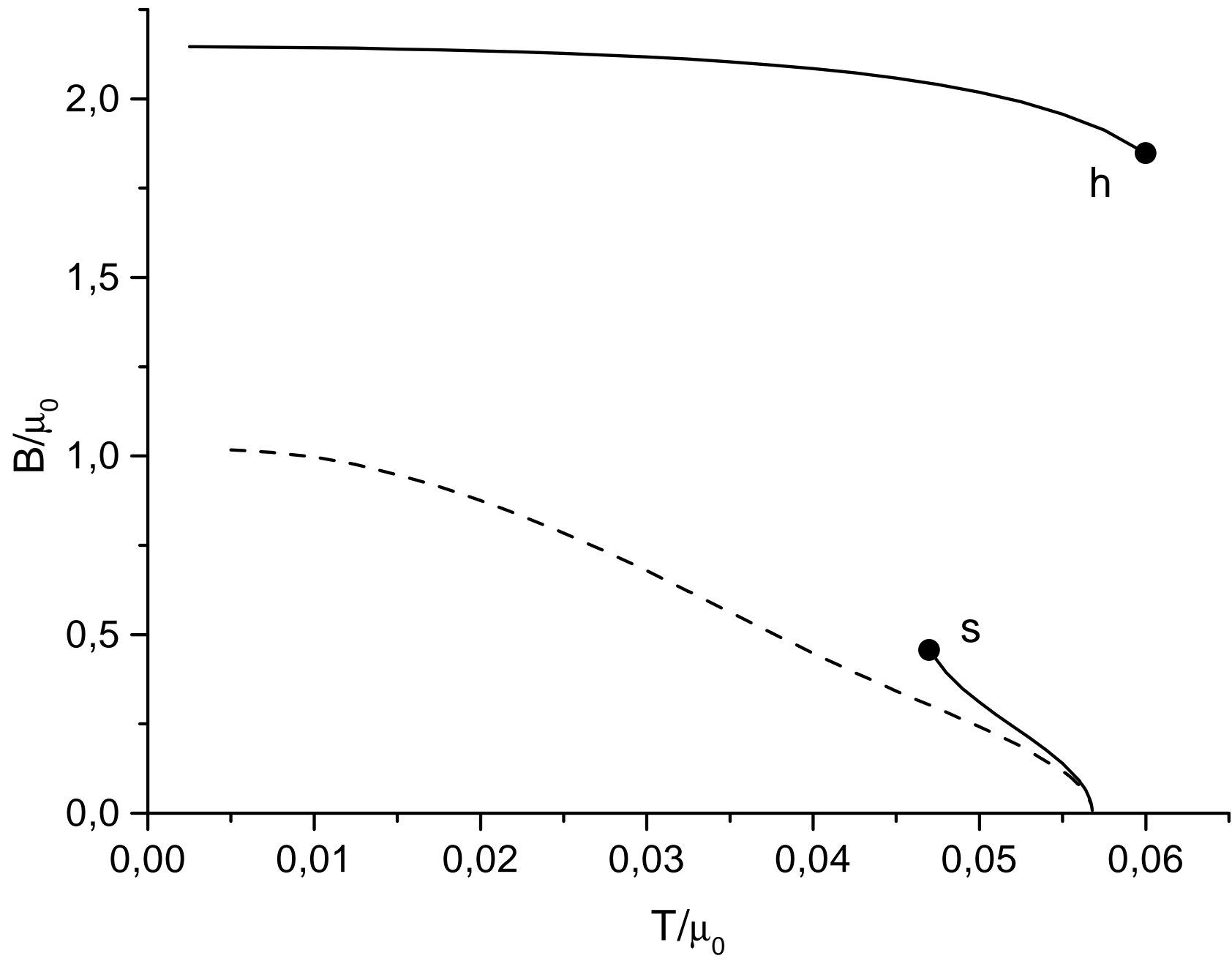
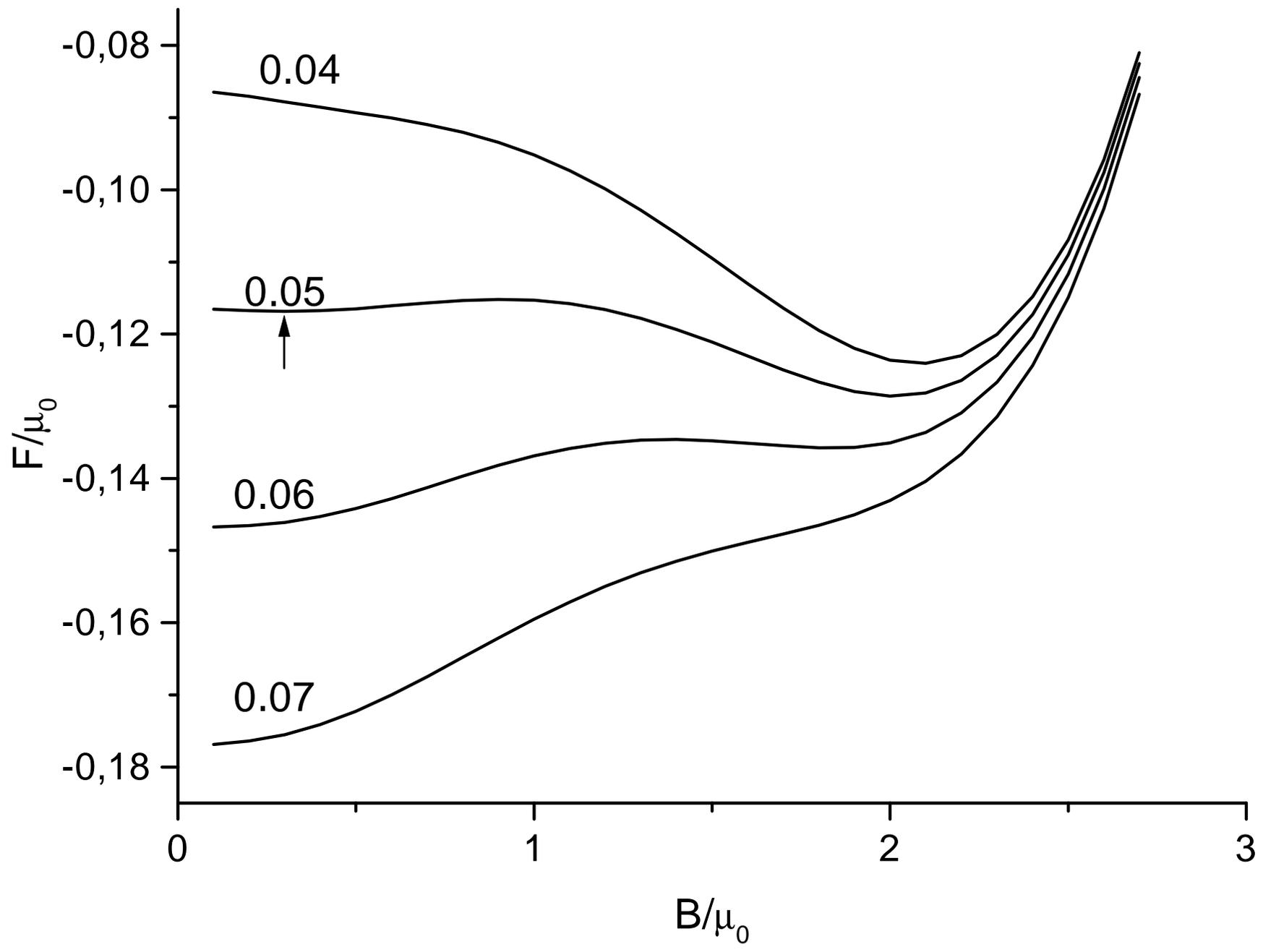
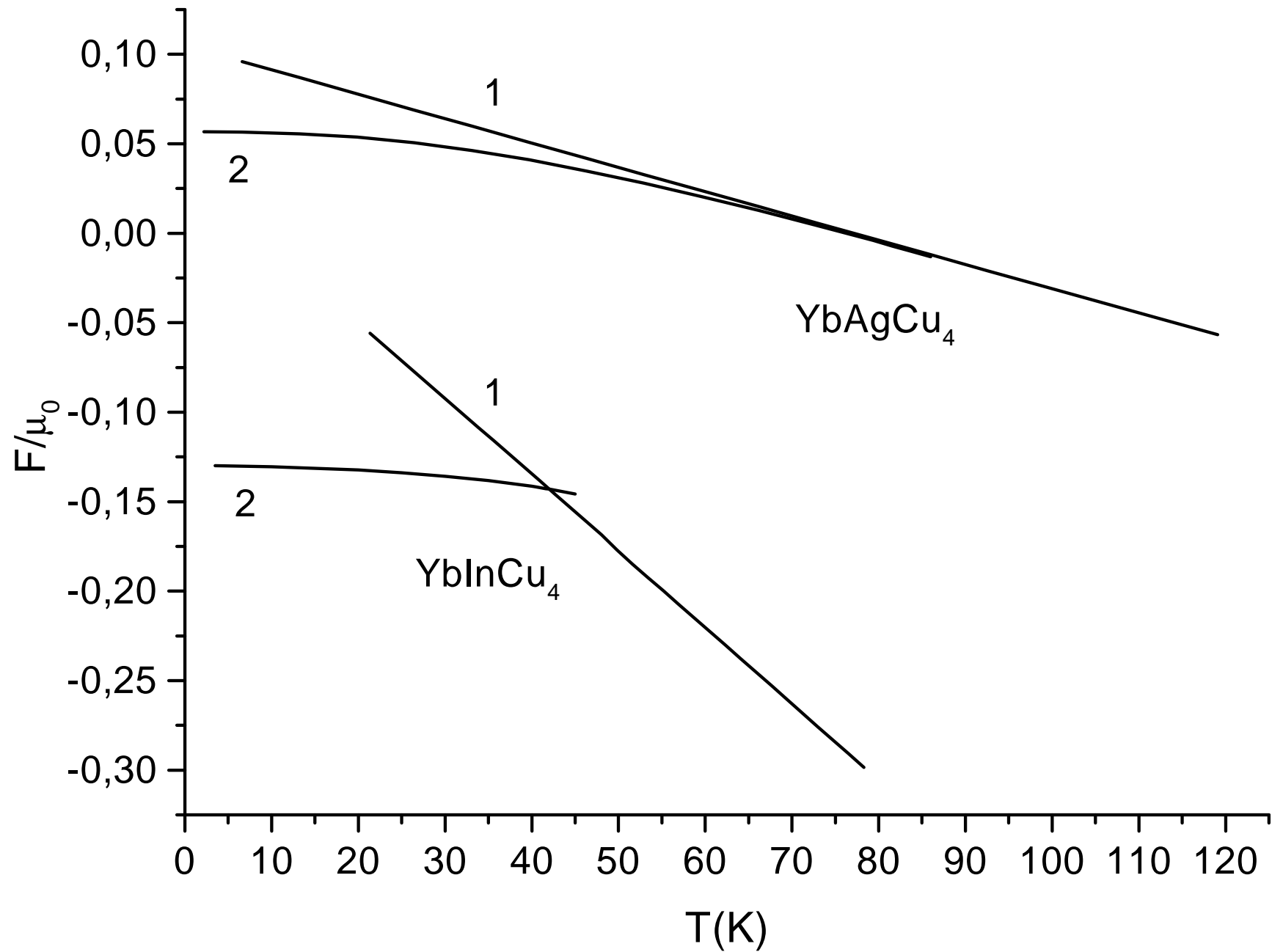
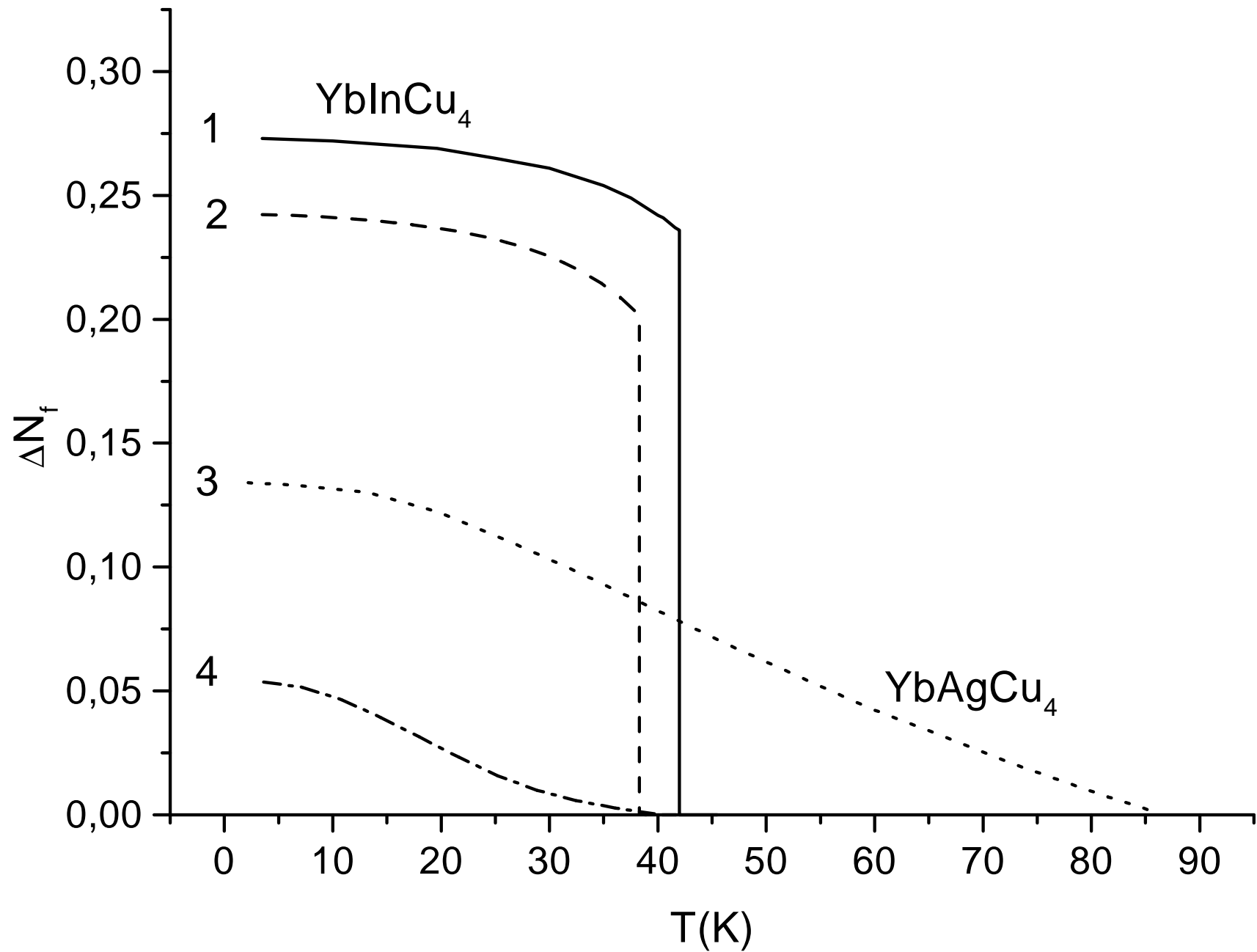
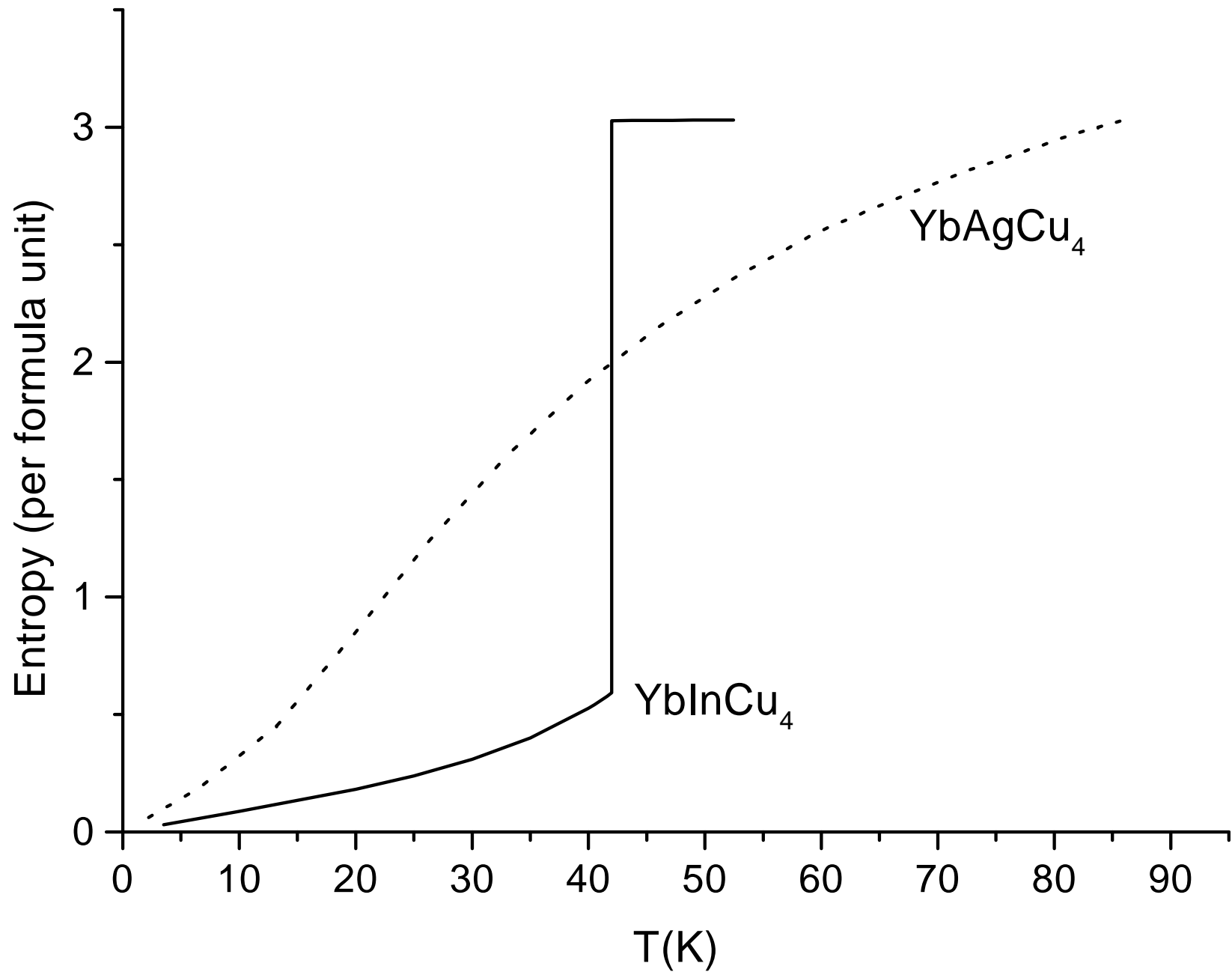
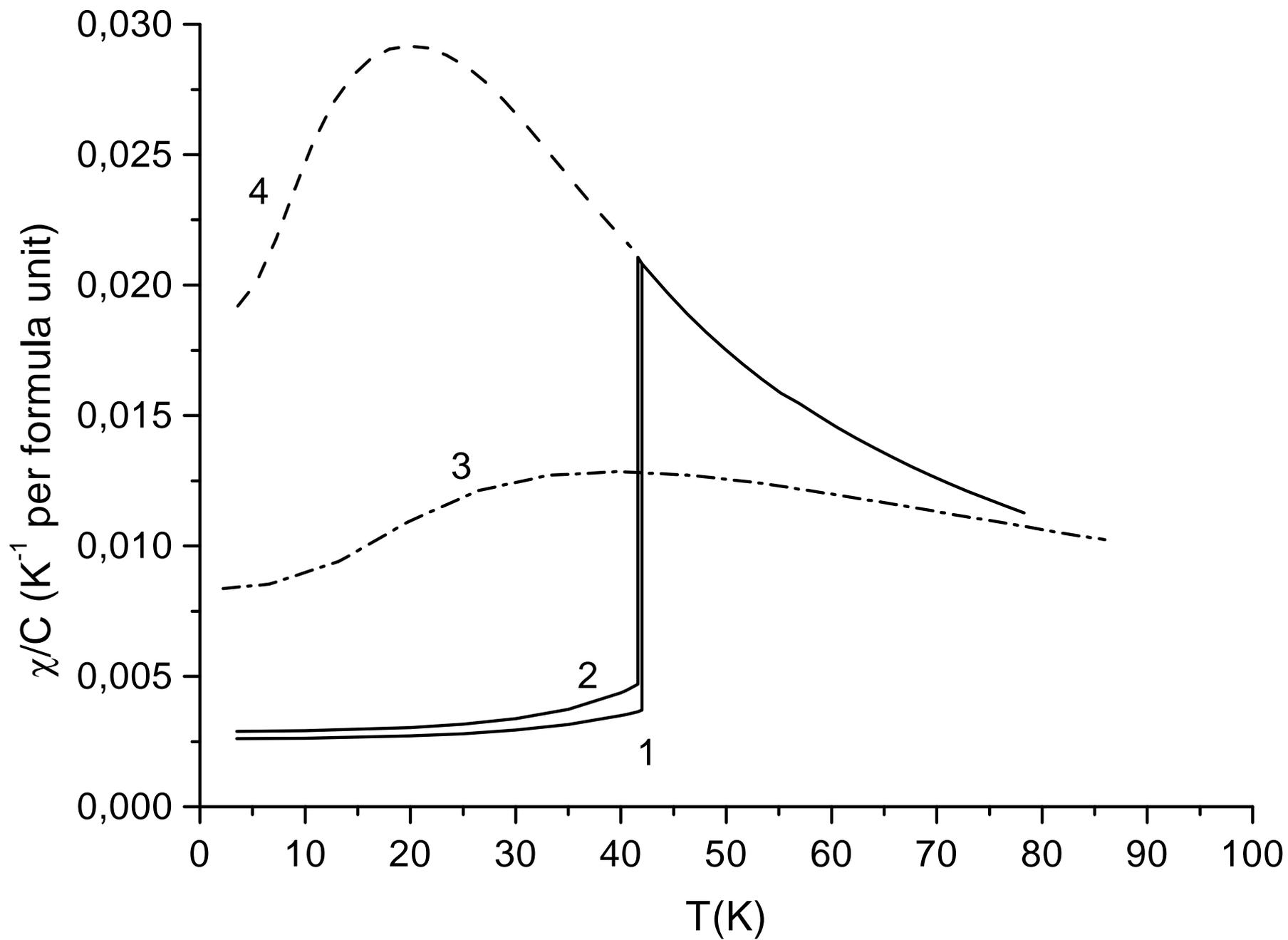
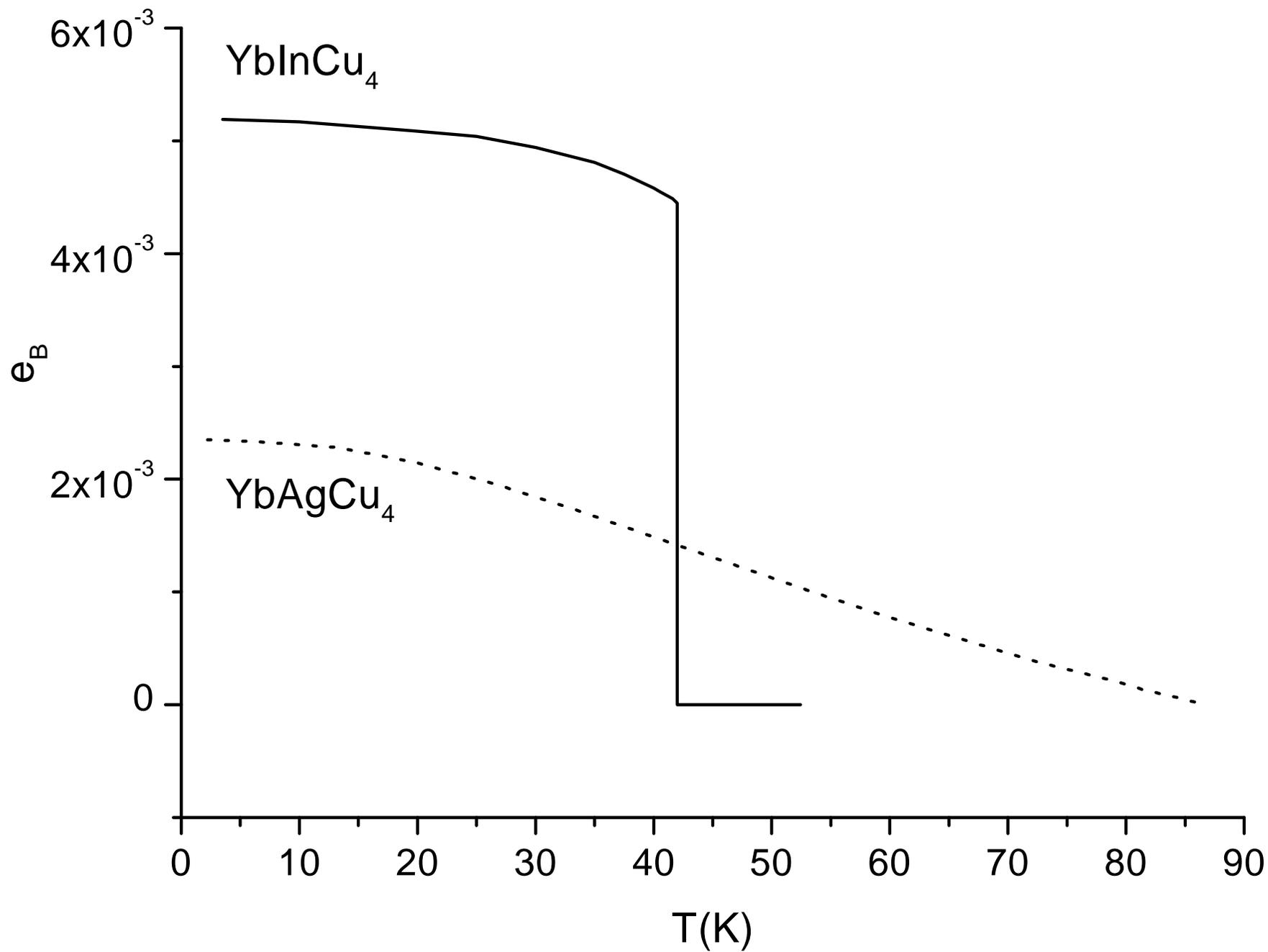
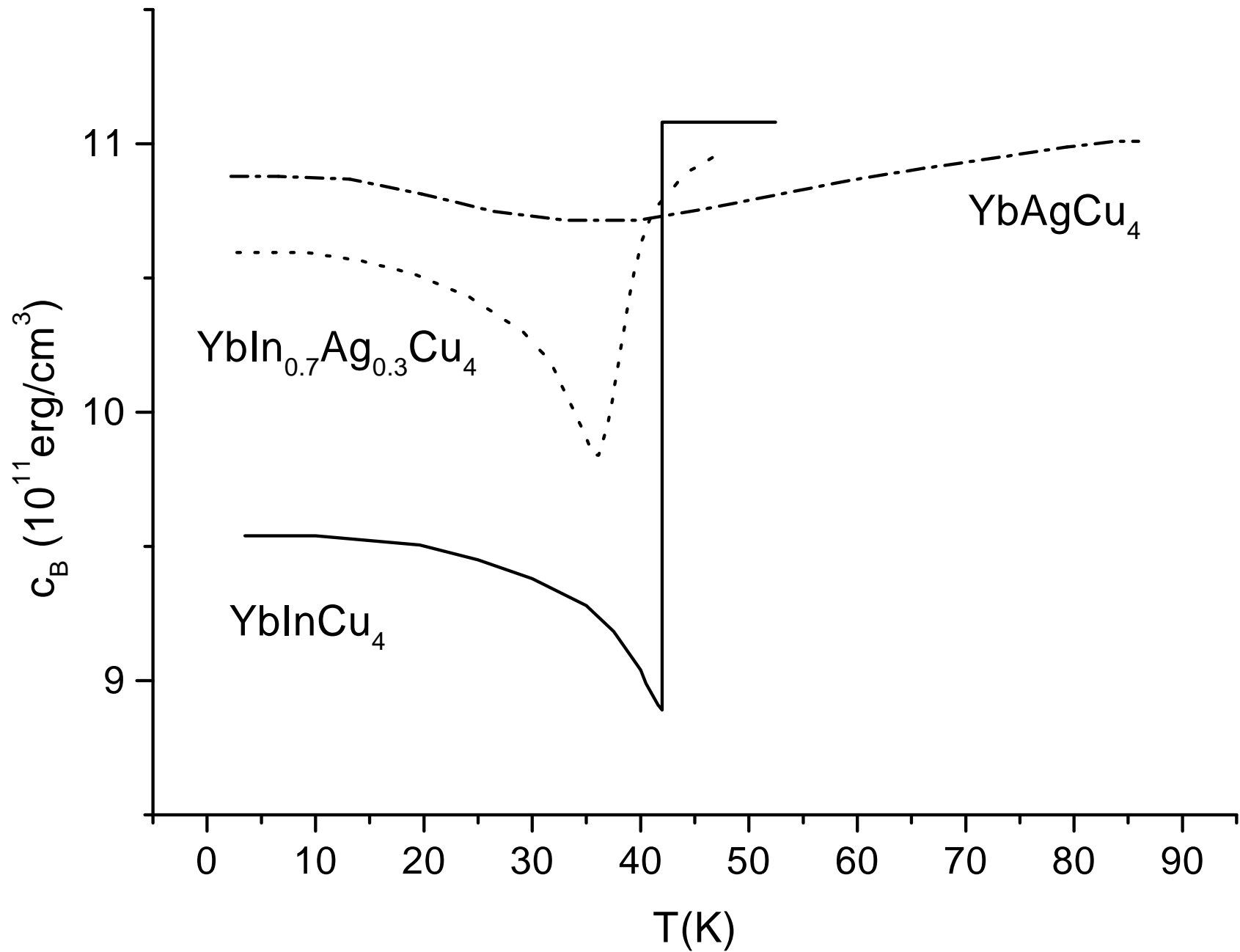
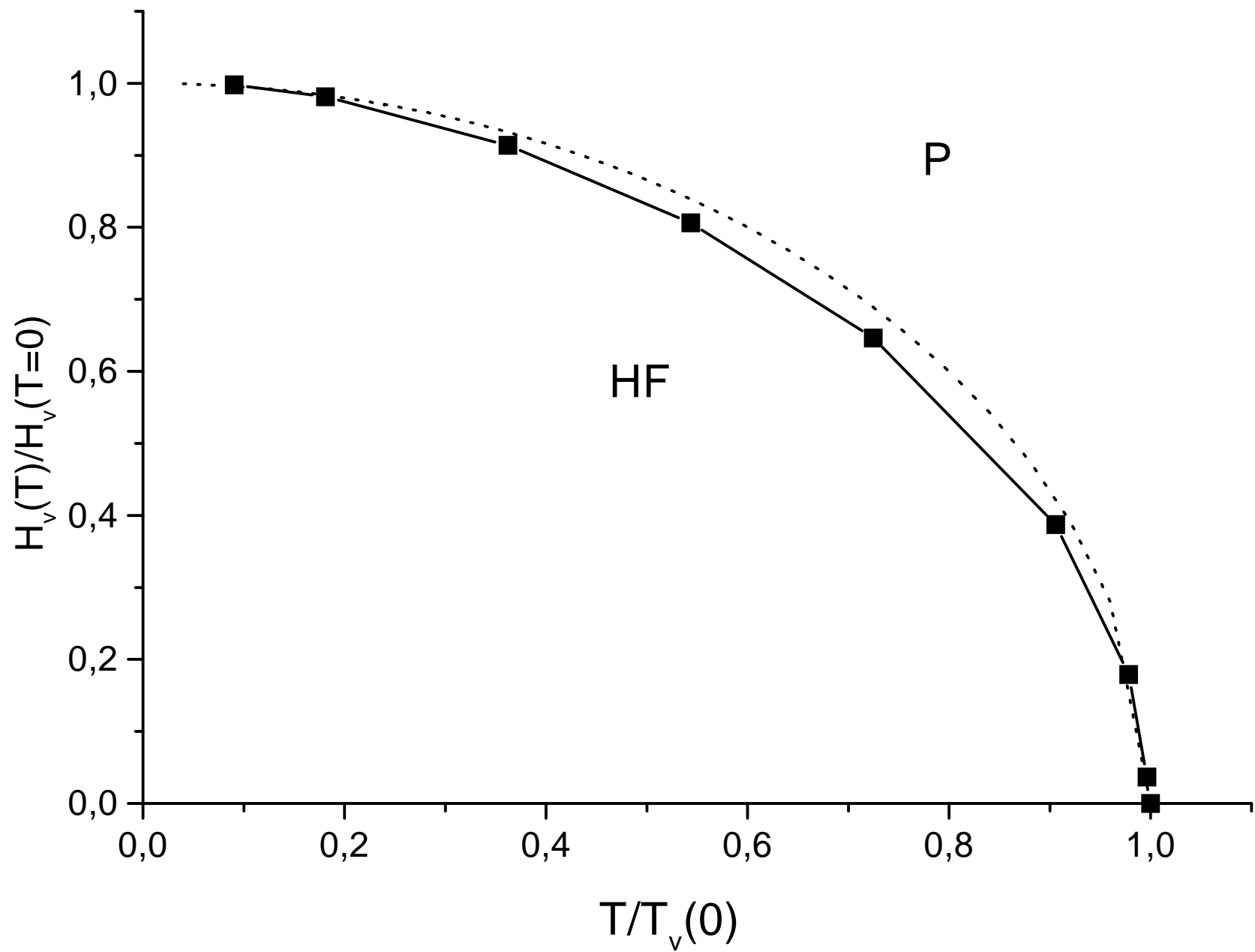
\end{document}